\documentclass[twocolumn,useAMS,usenatbib,revtex4-1,floatfix,superscriptaddress,nofootinbib]{mn2e}
\usepackage{times}
\usepackage{graphicx,natbib}
\usepackage{url}
\usepackage{color}
\usepackage{rotating}
\usepackage{subfigure}
\newcommand{\degree}{\hbox{$^\circ$}}

\newcommand{\fsky}{f_{\rm sky}}

\newcommand{\hinv}{h^{-1}}
\newcommand{\dwmean}{$\overline{\delta w/\sigma(w)}$}

\newcommand{\zphot}{z_{\rm p}}
\newcommand{\zspec}{z_{\rm s}}
\newcommand{\pofz}{$p(z)$}
\newcommand{\pofzw}{$p(z)_{{w}}$}

\newcommand{\Msun}{M_{\odot}}

\newcommand{\nwei}{N(z)_{\rm wei}}
\newcommand{\npz}{N(z)_{\rm p(z)}}

\newcommand{\nzsphot}{N(\zspec)^{\rm phot}}
\newcommand{\nzpts}{N(\zspec)^{\rm p37}}
\newcommand{\delpzs}{\Delta \pzs}
\newcommand{\chisq}{\Delta \chi^2_{\rm tot}}
\newcommand{\chisqm}{\Delta \chi^2_{\rm med}}
\newcommand{\degs}{\textrm{ deg}^2}
\newcommand{\degsb}{{\textrm{ \bf deg}^2}}

\newcommand{\acmins}{\textrm{ arcmin}^2}
\newcommand{\pzp}{P(z_p|z_s)}
\newcommand{\pzs}{P(z_s|z_p)}

\newcommand{\npoint}{N_{\rm point}}

\begin{document}

\title{Sample variance in photometric redshift calibration: cosmological biases and survey requirements}

\author[Cunha et al.]{Carlos E. Cunha$^{1,2}$\thanks{{\tt ccunha@stanford.edu}},
Dragan Huterer$^{1}$,
Michael T. Busha$^{2,3}$,
Risa H. Wechsler$^{2,4}$
\\
${}^{1}$Department of Physics, University of Michigan, 450 Church St, Ann Arbor, MI 48109-1040\\
${}^{2}$Kavli Institute for Particle Astrophysics and Cosmology 452 Lomita Mall, Stanford University, Stanford, CA, 94305\\
${}^{3}$Institute for Theoretical Physics, University of Zurich, 8057 Zurich, Switzerland\\
${}^{4}$Department of Physics, Stanford University, Stanford, CA, 94305,\\
SLAC National Accelerator Laboratory, 2575 Sand Hill Rd., MS 29, Menlo Park, CA, 94025
}

\date{\today}

\maketitle

\begin{abstract}
  We use N-body/photometric galaxy simulations to examine the impact of sample
  variance of spectroscopic redshift samples on the accuracy of photometric
  redshift (photo-z) determination and calibration of photo-z errors.  We
  estimate the biases in the cosmological parameter constraints from weak
  lensing and derive requirements on the spectroscopic follow-up for three
  different photo-z algorithms chosen to broadly span the range of algorithms
  available.  We find that sample variance is much more relevant for the
  photo-z error calibration than for photo-z training, implying that follow-up
  requirements are similar for different algorithms.  We demonstrate that the
  spectroscopic sample can be used for training of photo-zs and error
  calibration without incurring additional bias in the cosmological
  parameters. We provide a guide for observing proposals for
  the spectroscopic follow-up to ensure that redshift calibration biases do
  not dominate the cosmological parameter error budget.  For example, assuming
  optimistically (pessimistically) that the weak lensing shear measurements
  from the Dark Energy Survey could obtain $1-\sigma$ constraints on the dark
  energy equation of state $w$ of 0.035 (0.055), implies a follow-up
  requirement of 150 (40) patches of sky with a telescope such as Magellan,
  assuming a $1/8 \degs$ effective field of view and 400 galaxies per patch.
  Assuming (optimistically) a VVDS-like spectroscopic completeness with purely
  random failures, this could be accomplished with about 75 (20) nights of
  observation.  For more realistic assumptions regarding spectroscopic
  completeness, or with the presence of other sources of systematics not
  considered here, further degradations to dark energy constraints are
  possible.  We test several approaches for making the requirements less
    stringent.  For example, if the redshift distribution of the overall
    sample can be estimated by some other technique, e.g.\ cross-correlation,
    then follow-up requirements could be reduced by an order of magnitude.
\end{abstract}

\section{Introduction}\label{sec:intro}

One of the principal systematic errors affecting surveys that utilize the
large-scale structure to study dark energy is the quality of the photometric
redshifts (hereafter photo-zs).  Due to time and throughput constraints it is
costly and impractical to obtain spectroscopic redshifts for more than a small
fraction of galaxies.  Upcoming surveys such as the Dark Energy
Survey\footnote{\url{http://darkenergysurvey.org}} (DES),
PanStarrs\footnote{\url{http://pan-starrs.ifa.hawaii.edu}}, Hyper-Suprime Cam
survey\footnote{\url{http://oir.asiaa.sinica.edu.tw/hsc.php}} (HSC) and the
Large Synoptic Survey Telescope\footnote{\url{http://lsst.org}} (LSST) will
have to rely on the photo-zs in order to utilize the three-dimensional
information from the large number of galaxies observed in these surveys.
Without the redshift information, one loses the ability to perform weak
lensing tomography \cite{Hu_tomo}, and thus degrades the ability to measure
the temporal evolution of dark energy in the recent ($z\la 1$) history of the
universe \citep[for reviews,
  see][]{Bartelmann_Schneider,Huterer_thesis,Hu_tomo_2,Munshi_review,Hoekstra_Jain,Amara_Refregier_optimal,Huterer_GRG}.

Photo-z techniques use broad-band photometry, i.e. the measured flux through a
few bands, to estimate approximate galaxy redshifts.  Other observable
quantities (hereafter 'observables'), such as galaxy shape measures, can also
be used, but they typically have limited redshift information.  The
intrinsic uncertainty of photo-zs can contribute significantly to the error
  in the inferred cosmological parameters.

There are two broad, overlapping, categories of photo-z estimators.
Template-fitting algorithms \citep[e.g.][]{arn99,bol00,ben00,bud00,csa03,fel06}
assign photo-zs to a galaxy by finding the
template and redshift that best reproduce the observed fluxes.  Training set
methods \citep[e.g.][]{con95b,fir03,wad04,wan07,ger10}, on the 
other hand, use a spectroscopic sample to characterize a
relation between the photometric observables and the redshifts, which is then
applied to the full photometric sample.  The distinction between the two
categories is muddled because template-fitting methods can also use
spectroscopic redshifts to improve the fitting.  Conversely, training set
methods can be based on catalogs simulated using templates.
For reviews and comparison of methods see, e.g. \cite{hog98,koo99,hil10,abd11}.

Spectroscopic redshifts (hereafter spec-zs) play three important roles in
photo-z analysis.  First, as described above they improve the accuracy of
photo-z estimation.  Having accurate photo-zs is highly desirable for
cosmology, as photo-z errors inevitably smear the radial information
describing galaxy clustering.
Second, spectroscopic redshifts characterize the photo-z errors \citep[see
Ref.][for a review]{oya08b}.  With accurate error estimation, one
can remove or downweight the less reliable photo-zs, decreasing their
impact on the cosmological analysis.  Third, spec-zs characterize the
uncertainties in the photo-z error distribution ('error in the
error'), which is a key quantity that needs to be accurately known.
In particular, even if photo-zs are not exceptionally accurate and
there are regions of badly misestimated redshift (i.e.\ the
'catastrophic errors'), one can still recover the cosmological
information provided the bias, scatter, and ideally the full
distribution in the $\zphot-\zspec$ plane, are accurately calibrated
using the subset of galaxies with spectroscopic information.

The requirements on spectroscopic samples due to the three
requirements mentioned are not independent, but have been treated as
such in the past.  For example, the amount of spectroscopic follow-up
required for calibration depends on the intrinsic accuracy of the
photo-zs \citep{ma06,hut06,Bernstein_Ma} and on the identification of
regions with unreliable photo-zs \citep{BH10,Sun,Hearin}, but the
ability to do both of these is strongly dependent on the use of
spec-zs for training and error estimation.  At this point, the careful
reader may wonder: can the same spectroscopic sample be used for
photo-z training, error estimation and calibration without
significantly biasing cosmological results?  Yes, it turns out, as we
will show in this paper.

Obtaining spectra for thousands of galaxies needed for photo-z studies
is a very difficult task, which complicates their use in photo-z
studies.  Spectroscopic surveys can be far from a representative
sub-sample of the photometric sample for five principal reasons:
\begin{itemize}
\item {\it Shot noise:} Spectroscopic samples to the depth required are quite 
small, hence Poisson fluctuations due to the finite number of galaxies are
significant.

\item {\it Sample variance:} Spectroscopic surveys designed to reach the
  magnitude limits of the upcoming photometric surveys typically have very
  small angular apertures, much smaller than fluctuations introduced by
  large-scale clustering of galaxies.  The fluctuations due to sample variance
  can be an order of magnitude larger than shot-noise fluctuations for samples
  of around $1 \degs$ \citep[see e.g.][and Fig.~\ref{fig:nzspec}]{van06}.

\item {\it Type incompleteness:} Strength of spectral features vary significantly 
with galaxy type. In addition, the wavelength coverage of most spectrographs is not 
sufficient to detect some of the main features through the full redshift range of interest.

\item {\it Incorrect redshifts:} Line misidentification can yield incorrect redshifts. 
The number of incorrect spectroscopic redshifts can be reduced -- by keeping 
only the most reliable galaxies --  at the cost of increasing the incompleteness.

\item {\it Sample variance in observing conditions:} Variations in imaging conditions 
(e.g. seeing and photometric quality) during a survey imprint an angular dependence 
to the survey depth and completeness.

\end{itemize}

Past papers on the effects of photometric redshift errors on dark energy
constraints
\citep{ma06,hut06,Amara_Refregier_optimal,Abdalla08,Bernstein_Ma,Kitching_sys,bor10,Hearin}
have studied in detail the distribution of photometric redshifts (more
specifically, the full probability density function $P(z_p|z_s)$). Some of
these works have extended the analysis to estimate the number of spectra
required in order to calibrate the photo-zs. However, in essentially all cases
the requirements on the spectroscopic sample have only assumed shot noise,
i.e.\ that the accuracy of the photo-z bias and error in some redshift bin
labeled by $\mu$ is equal to $ \Delta z_{\rm
  bias}(z_{\mu})=\sigma_z(z_{\mu})\sqrt{1 / N_{\rm spec}^{\mu}} $ and
$\Delta\sigma_z(z_{\mu}) = \sigma_z(z_{\mu}) \sqrt{2 /N_{\rm spec}^{\mu}}$,
where $N_{\rm spec}^{\mu}$ is the size of the spectroscopic follow-up sample
in that bin \citep[see Eq.~(18) in Ref.][]{ma06}.

Sample variance was taken into account in spectroscopic follow-up requirements
in \cite{van06,ish05}; however,
they only considered the overall redshift
distribution of the source sample and did not include photometric redshifts in
the simulations \citep[see also][for a related discussion]{bor10}. Requirements
on spectrograph design in order to minimize spectroscopic failures were
investigated in \cite{jou09}, with emphasis on designing spectrographs to
calibrate redshifts for space-based missions. Finally, the effects of sample
variance in observing conditions was investigated by \cite{nak11} using SDSS
imaging and spectroscopic redshifts from several surveys overlapping the SDSS.
That paper found that atypical imaging conditions in the spectroscopic fields
can lead to biases in galaxy-galaxy lensing analysis, but fortunately concluded
that this type of bias can be at least partly corrected 
\citep[see also][for a related discussion]{she11}.

The main goals of this paper are to study the impact of sample variance in
spectroscopic samples to the training of photo-zs, error estimation and error
calibration, and to assess implications for cosmological constraints from weak
lensing tomography analyses.  The paper is organized as follows.  In
Sec.\ \ref{sec:algs} we describe the photo-z algorithms we use in our tests.
In Sec.\ \ref{sec:data} we describe our construction of the different simulated
samples used.  We detail the procedure of estimating biases in cosmological
constraints from the weak lensing tomography in Sec.\ \ref{sec:wl}.  Results
are given in Sec.\ \ref{sec:results} with a discussion of potential
improvements in Sec.\ \ref{sec:improv}. We provide a guide for determining
spectroscopic observational requirements in Sec.\ \ref{sec:guide} and 
present our conclusions in Sec.\ \ref{sec:concl}.  The construction of the
simulations is described in Appendix \ref{app:sims}.

\section{Photo-z algorithms}\label{sec:algs}

We consider three different photo-z algorithms that broadly span the space of
possibilities.  Namely, we use a basic template-fitting code without any
priors, a training set fitting method, and a training set method that does not
perform a fit, but uses the local density in the neighborhood of an object to
derive a redshift probability distribution.
We briefly describe each below.

\subsection{Template-fitting redshift estimators}\label{sec:templ}

Template-fitting estimators derive photometric redshift estimates by comparing
the observed colors of galaxies to colors predicted from a library of galaxy
spectral energy distributions.  We use the publicly available   {\it LePhare}
photo-z code\footnote{\url{
  http://www.cfht.hawaii.edu/~arnouts/LEPHARE/lephare.html}} \citep{arn99,ilb06}
as our template-fitting estimator.  We chose the extended CWW template
library \citep{col80} because it yielded the best photo-z's for our simulation.

We purposefully ignore all priors for reasons that we now describe.  There are
essentially two classes of priors, those derived from completely
different surveys, and those based on targeted follow-ups of a subsample of
the survey for which photo-z's are desired.  The use of the latter makes
template-fitting results quite similar to the training set methods, and would make
the template-fitting code subject to a training procedure which would be
affected by the sample variance.  The use of the former could reduce some
outliers, but would also complicate the interpretation of the results;  there
are several choices of external priors, and if the selection of the
sample used to determine the priors is different from that of the survey at
hand then
redshifts could be biased \citep[see, e.g.][]{abr11}. 
As we shall see, the photo-z quality is not a
dominant factor in our analysis, and a more thorough experimentation of the 
template-fitting algorithms is not expected to affect conclusions.

\subsection{Nearest-neighbor redshift probability estimators}\label{sec:nnpz}

\subsubsection{Weights }

In this subsection, we briefly review the weighting method\footnote{The
  weights, \pofz\ and polynomial codes are available at \url{
    http://kobayashi.physics.lsa.umich.edu/~ccunha/nearest/}. 
The codes can also be obtained in the git repository {probwts} at \url{http://github.com}} of \cite{lim08},
which is required for computing redshift probabilities, henceforth \pofz.  We
define the weight, $w$, 
of a galaxy in the spectroscopic training set as the
normalized ratio of the density of galaxies in the photometric sample to the
density of training-set galaxies around the given galaxy.  These densities are
calculated in a local neighborhood in the space of photometric observables,
e.g.\ multi-band magnitudes.  In this case, the DES {\it griz} magnitudes are
our observables.  The hypervolume used to estimate the density is set here to
be the Euclidean distance of the galaxy to its $N^{\rm th}$ nearest-neighbor in
the training set.  We set $N=50$ for the \pofz\ estimate.  Smaller $N$ lead to
less broad \pofz s and better reconstruction of the overall redshift
distribution  at the cost of increased shot-noise in individual \pofz s.  If
one does not care about individual \pofz s then we recommend choosing a 
smaller N; the optimal choice will depend on the training set size.  
The bias analysis is not sensitive to the choice of $N$.

The weights can be used to estimate the redshift distribution of the
photometric sample using
\begin{equation}  
\nwei   = \sum_{\beta=1}^{N_{\rm T,tot}} w_\beta N(z_1<z_\beta<z_2)_{\rm T},
\label{eqn:Nzest}
\end{equation}
where the weighted sum is over all galaxies in the training set. \citet{lim08}
and \citet{cun09} show that this provides a nearly unbiased estimate of
the redshift distribution of the photometric sample, $N(z)_{\rm P}$, provided
the differences in the selection of the training and photometric samples are
solely done in the observable quantities used to calculate the weights.

\subsubsection{Probability density \pofzw}

To estimate the redshift error distribution for each galaxy, \pofzw,
we adopt the method of \citet{cun09}.
We use the subscript $w$ to differentiate between our particular estimator and 
the general concept for redshift probability distributions.
The \pofzw\ for a given object in the
photometric sample is simply the redshift distribution of the $N$ (in this case
50) nearest-neighbors in the {\it training} set 

\begin{equation}
{p}(z)_w = \sum_{\beta=1}^{N} w_\beta \delta(z-z_\beta)~.
\label{eqn:pzest}
\end{equation}
We estimate \pofzw\ in 20 redshift bins between $z=0$ and $1.35$.  

We can also construct a new estimator for the number of galaxies $N(z)_{\rm P}$
by summing the $p(z)_w$ distributions for all galaxies in the photometric
sample
\begin{equation}
\npz = \sum_{i=1}^{N_{\rm P,tot}}p_i(z)_w~.
\label{eqn:Nhat2}
\end{equation}
This estimator becomes identical to that of Eq.~(\ref{eqn:Nzest}) in the limit
of very large training sets.  For training sets smaller than tens of thousands
of galaxies, one can improve the \pofzw\   estimate by
multiplying each \pofzw\ by the ratio of $\nwei$ to $\npz$.

We note that several public photo-z codes exist that can output \pofz s, e.g.,
the template-fitting codes {\it Le Phare} \citep{arn99,ilb06}, {\it ZEBRA}
\citep{fel06}, {\it BPZ} \citep{coe06}, and the training-set based {\it
  ArborZ} \citep{ger10}.  We do not expect qualitative differences in our
conclusions from using the above methods because, as we will show, sample
variance affects mostly spectroscopic properties, not photometric.

\subsection{Nearest-neighbor polynomial fitting redshift estimators}\label{sec:nnp}

For each galaxy in the photometric sample, the nearest-neighbor
polynomial fitting algorithm (NNP) uses the $N$ nearest neighboring
galaxies with spectra (i.e. in the training set) to fit a low-order
polynomial relation between the redshift and the observable quantities
(e.g.\ colors and magnitudes).  It then applies this function to the
observables of the galaxy in the photometric sample and assigns it a
redshift.  We use a second-order polynomial in this study and check
that a first-order polynomial does not change results by more than a
few percent.  The NNP method was introduced by \citet{oya08a} and
produces photo-z's that are very similar to the neural networks.  We
chose the NNP here because it is very fast compared to other codes for
photometric samples with up to a few million objects in size.  In
addition, we can directly compare the results of the NNP photo-z's
with the \pofzw\ since both are based on the same set of training-set
galaxies.  As with the \pofzw\ method, the choice of which $N$
nearest-neighbors are to be used does not affect results
significantly, provided there are enough galaxies to characterize the
coefficients of the polynomial fit and avoid over-fitting.  For a
second-order polynomial with 4 observables, we find that $N=100$ is a
good comprise between retaining locality of color information and
stability of the fit.  Results presented here use a slightly more
agressive $N=80$, but this does not affect the bias results
meaningfully.

\section{Simulated Data}\label{sec:data}

\subsection{Selection}

We use a cosmological simulation, populated with galaxies and their
photometric properties, fully described in Appendix \ref{app:sims}.
The simulation consists of a $220 \degs$ photometric survey in the
grizY DES bands with $10\sigma$ magnitude limits of $[24.6, 24.1,
24.4, 23.8, 21.3]$.  For this study, we disregard the $Y$-band since
we find it does not improve the photo-z's.  We select only galaxies
with $i<24$ which are also detected (to $5\sigma$) in the grz bands.
The original catalog contains 13,550,386 galaxies, and after the cuts
we are left with $N_{\rm data}$=10,780,625 galaxies.  To speed up the
training and calibration of the photo-z's, we pick a random subsample
of about $N_{\rm phot} =4,000,000$ galaxies to be our photometric
sample.

\subsection{Training and calibration samples}

We construct our spectroscopic training and calibration samples by splitting
the simulation output into several sets of $N\times N$ patches of equal area,
with each patch being nearly square in shape.  When comparing the different
photo-z algorithms we use three binning schemes, setting $N=6$, 15, and 30,
which corresponds, roughly, to patches of area 6, 1, and 0.25 $\degs$
respectively.  Because spectroscopic surveys are far from complete, in a sense
that they include spectra of only a subset of all photometrically discovered
galaxies, we randomly pick a subsample from each patch.  Unless stated
otherwise, we simulate $25\%$ random completeness, that is, we use a Monte Carlo
approach to downsample by drawing a random number between 0 and 1 for each galaxy 
and selecting the galaxies for which the number is less than 0.25.
The mean number of galaxies per pixel available for training and calibration is 
about 74,865, 11,978, and 2,995 for the 6, 1 and 0.25 $\degs$ pixel sets.  
We refer to the sample
created by splitting the data in angular patches as the {\it large-scale structure
(LSS) samples}.

For each set of LSS samples, we generate what we call the {\it
  random-equivalent samples}.  The random-equivalent samples are sets
of random samples drawn from the full survey but with size similar to
the LSS sample patches.  For example, the random equivalent patches of
the 1 $\degs$ LSS patches are generated as follows.  There are 225
patches in the 1 $\degs$ case.  The random equivalent patches are
generated by performing random draws of galaxies from the full data
set to generate a new set of 225 patches; each such (random
equivalent) patch is generated by including every galaxy from the
original catalog with the probability $N_{\rm patch}/N_{\rm gal}$,
where $N_{\rm patch}$ is the average number per patch (eg.\ 11,978 in
the 1$\degs$ case), while $N_{\rm gal}$ is the total number of
galaxies in the simulation.  This yields 225 samples that have the
same average number of galaxies per patch as the LSS patches.

As discussed in the Introduction, in real spectroscopic surveys the
incompleteness is caused not only by random sub-selection, but also the
inability to get spectra for some galaxies.  These spectroscopic failures can
lead to biases in the training and calibration and we shall explore them in a
follow-up paper.  Throughout, we use the same set of patches for both training
and calibration.  In Sec. \ref{sec:results}, we show that this does not add
appreciable error to the cosmological constraints.

\section{Weak Lensing bias} \label{sec:wl}

We wish to quantify how much sample variance due to the LSS
contributes to errors in weak lensing shear, and thus errors in the
derived cosmological parameter constraints.  For simplicity, we only
study the shear-shear correlations, and not the related shear-galaxy
and galaxy-galaxy power spectra.  The observable quantity we consider
is the convergence power spectrum
\begin{equation}
C^{\kappa}_{ij}(\ell)=P_{ij}^{\kappa}(\ell) + 
\delta_{ij} {\langle \gamma_{\rm int}^2\rangle \over \bar{n}_i},
\label{eq:C_obs}
\end{equation}
where $\langle\gamma_{\rm int}^2\rangle^{1/2}$ is the rms intrinsic
ellipticity in each component, $\bar{n}_i$ 
is the average number of galaxies in the $i$th redshift bin per steradian, and $\ell$ is the
multipole that corresponds to structures subtending the angle $\theta =
180\degree/\ell$.  For simplicity, we drop the superscripts $\kappa$ below. 
For most of this work we take $\langle\gamma_{\rm int}^2\rangle^{1/2}=0.16$,
which yields very stringent follow-up requirements. We discuss the impact 
of this choice in Sec. \ref{sec:disce}.

We closely follow the formalism of \citet{BH10} (hereafter BH10), where the
photometric redshift errors are algebraically propagated into the biases in
the shear power spectra. These biases in the shear spectra can then be
straightforwardly propagated into the biases in the cosmological parameters.
We now review briefly this approach.

Let us assume a survey with the (true) distribution of source galaxies in
redshift $n_S(z)$, divided into $B$ bins in redshift. Let us define the
following terms
\begin{itemize}
\item {\em Leakage} $P(z_p|z_s)$ (or $l_{sp}$ in BH10 terminology): fraction
  of objects from a given spectroscopic bin that are placed into an incorrect
  (non-corresponding) photometric bin.

\item {\em Contamination} $P(z_s|z_p)$ (or $c_{sp}$ in BH10 terminology): fraction of galaxies
  in a given photometric bin that come from a non-corresponding spectroscopic
  bin.
\end{itemize}

When specified for each tomographic bin, these two quantities contain the same
information.  Note in particular that the two quantities satisfy the
integrability conditions
\begin{eqnarray}
\int P(z_p|z_s)dz_p &\equiv& \sum_p l_{sp} = 1\\[0.2cm]
\int P(z_s|z_p)dz_s &\equiv& \sum_s c_{sp} = 1.
\end{eqnarray}

A fraction $l_{sp}$ of galaxies in some
spectroscopic-redshift bin $n_s$ ``leak'' into some photo-z bin $n_p$, so that
$l_{sp}$ is the fractional perturbation in the spectroscopic bin, while the
contamination $c_{sp}$ is the fractional perturbation in the photometric bin.
The two quantities can be related via
\begin{equation}
c_{sp} = \frac{N_s}{N_p}\, l_{sp} 
\label{eq:contamination}
\end{equation}
where $N_s$ and $N_p$ are the absolute galaxy numbers in the spectroscopic
and photometric bin respectively. Then
\begin{eqnarray}
n_s &\rightarrow & n_s\\[0.1cm]
n_p &\rightarrow & (1-c_{sp})\,n_p + c_{sp}\,n_s
\label{eq:n_bias}
\end{eqnarray}
and the photometric bin normalized number density is affected
(i.e. biased) by photo-z catastrophic errors.
The effect on the cross power spectra is then \cite{BH10}
\begin{eqnarray}
C_{pp} &\rightarrow &(1-c_{sp})^2C_{pp} +2c_{sp} (1-c_{sp})C_{sp} + c_{sp}^2 C_{ss}
           \nonumber\\[0.1cm]
C_{m p} &\rightarrow &(1-c_{sp})C_{mp} +c_{sp}\, C_{ms} 
           \qquad (m< p)
	   \label{eq:C_bias}\\[0.1cm]
C_{p n} &\rightarrow &(1-c_{sp})C_{pn} +c_{sp}\, C_{sn} 
           \hspace{0.9cm} (p < n)\nonumber \\[0.1cm]
C_{mn} &\rightarrow &C_{mn}\qquad\qquad\qquad\qquad\qquad  ({\rm otherwise})
	   \nonumber
\end{eqnarray}
(since the cross power spectra are symmetrical with respect to the interchange
of indices, we only consider the biases in power spectra $C_{ij}$ with $i\leq j$).  
Note that these equations are exact for a fixed contamination coefficient $c_{sp}$. 

The bias in the observable power spectra is the rhs-lhs difference in the
above equations\footnote{We have checked that the quadratic terms in $c_{sp}$
  are unimportant, but we include them in any case.}.  The cumulative result
due to all contaminations in the survey (or, $P(z_s|z_p)$ values for each
$z_s$ and $z_p$ binned value) can be obtained by the appropriate sum
\begin{eqnarray}
\delta C_{pp} &=&\sum_s (-2c_{sp} + c_{sp}^2)C_{pp} +2c_{sp} (1-c_{sp})C_{sp} + c_{sp}^2 C_{ss}
           \nonumber\\[0.1cm]
\delta C_{mp} & =&\sum_s \left (-c_{sp}C_{mp} +c_{sp}\, C_{ms}\right )
  \label{eq:delta_Cmp}\\[0.1cm]
\delta C_{pn} & =&\sum_s \left (-c_{sp}C_{pn} +c_{sp}\, C_{s n}\right )  \nonumber
\end{eqnarray}
for each pair of indices $(m, p)$, where the second and third line assume
$m<p$ and $p<n$, respectively.

The
bias in cosmological parameters is given by using the standard linearized
formula \citep{Knox_Scocc_Dod,Huterer_Turner},  summing over each pair of
contaminations $(s, p)$
\begin{equation}
\delta p_i \approx \sum_{j}(F^{-1})_{ij} \sum_{\alpha\beta} {\partial \bar
  C_\alpha \over \partial p_j} ({\rm Cov}^{-1})_{\alpha\beta}\,\delta C_\beta ,
\label{bias1}
\end{equation}
where $F$ is the Fisher matrix and ${\rm \bf Cov}$ is the covariance of shear
power spectra (see just below for definitions). This formula is accurate when
the biases are 'small', that is, when the biases in the cosmological
parameters are much smaller than statistical errors in them, or $\delta p_i
\ll (F^{-1})_{ii}^{1/2}$.  Here $i$ and $j$ label cosmological parameters, and
$\alpha$ and $\beta$ each denote a {\it pair} of tomographic bins,
i.e.\ $\alpha, \beta=1, 2, \ldots, B(B+1)/2$, where recall $B$ is the number
of tomographic redshift bins. 
To connect to the $C_{mn}$ notation in
Eq.~(\ref{eq:C_bias}), for example, we have $\beta = mB + n$.

We calculate the Fisher matrix $F$ assuming perfect redshifts, and following
the procedure used in many other papers \citep[e.g.][]{MMG}. 
The weak lensing Fisher matrix is then given by
\begin{equation}
F^{\rm WL}_{ij} = \sum_{\ell} \,{\partial {\bf C}\over \partial p_i}\,
{\bf Cov}^{-1}\,
{\partial {\bf C}\over \partial p_j},\label{eq:latter_F}
\end{equation}
where $p_i$ are the cosmological parameters and ${\bf Cov}^{-1}$ is
the inverse of the covariance matrix between the observed power spectra whose
elements are given by
\begin{eqnarray}
{\rm Cov}\left [C_{ij}(\ell), C_{kl}(\ell')\right ] &=& 
{\delta_{\ell \ell'}\over (2\ell+1)\,f_{\rm sky}\,\Delta \ell} \times \\
&&\left [ C_{ik}(\ell) C_{jl}(\ell) + 
  C_{il}(\ell) C_{jk}(\ell)\right ]. \nonumber 
\label{eq:Cov}
\end{eqnarray}
The fiducial weak lensing survey corresponds to expectations from the Dark
Energy Survey, and assumes 5000 square degrees (corresponding to $\fsky\simeq
0.12$) with tomographic measurements in $B=20$ uniformly wide redshift bins
extending out to $z_{\rm max}=1.35$. The effective source galaxy density is 12 galaxies
per square arcminute, while the maximum multipole considered in the
convergence power spectrum is $\ell_{\rm max}=1500$. The radial distribution
of galaxies, required to determine tomographic normalized number densities
$n_i$ in Eq.~(\ref{eq:C_obs}), is determined from the simulations and shown 
in Fig.~\ref{fig:nzspec}. 

We consider a standard set of six cosmological parameters with the following
fiducial values: matter density relative to critical $\Omega_M=0.25$, equation
of state parameter $w=-1$, physical baryon fraction $\Omega_B h^2=0.023$,
physical matter fraction $\Omega_M h^2=0.1225$ (corresponding to the scaled
Hubble constant $h=0.7$), spectral index $n=0.96$, and amplitude of the matter
power spectrum $\ln A$ where $A=2.3\times 10^{-9}$ (corresponding to
$\sigma_8=0.8$).  Finally, we add the information expected from the Planck
survey given by the Planck Fisher matrix (W.\ Hu, private communication). The
total Fisher matrix we use is thus
\begin{equation}
F = F^{\rm WL}+F^{\rm Planck}.
\label{eq:Fisher}
\end{equation}
The fiducial constraint on the equation of state of dark energy assuming
perfect knowledge of photometric redshifts is $\sigma(w)=0.035$.

Our goal is to estimate the biases in the cosmological parameters due to
imperfect knowledge of the photometric redshifts. In particular, the relevant
photo-z error will be the difference between the inferred $P(z_s|z_p)$
distribution for the calibration (or, training) set and that for the actual
survey.  Therefore, we define 
\begin{eqnarray}
\delta C_\beta &=& C_\beta^{\rm train} - C_\beta^{\rm phot}\noindent\\[0.2cm]
&=& \delta C_\beta^{\rm train} - \delta C_\beta^{\rm phot}
\end{eqnarray}
where the second line trivially follows given that the true, underlying power
spectra are the same for the training and photometric galaxies. All of the
shear power spectra biases $\delta C$ can straightforwardly be evaluated from
Eq.~(\ref{eq:delta_Cmp}) by using the contamination coefficients for the
training and photometric fields, respectively.  Therefore, the effective error
in the power spectra is equal to the difference in the biases of the training
set spectra (our {\it estimates} of the biases in the observable quantities) and the
photometric set spectra (the actual biases in the observables).

\section{Results}\label{sec:results}

We present our results in this section.  In Sec. \ref{sec:varvar} we compare
the effects of sample variance on the spectroscopic redshifts and the
photometric observables, concluding that the effects on the redshifts are
dominant.  We then discuss the impact of sample variance on photo-z training
in Sec.~\ref{sec:vartra}, finding that the effect on the photo-z scatter
statistics is negligible, but that it does introduce variability in the
estimate of the overall redshift distribution.  The effect is much smaller for
photo-z methods that use a fitting-function, such as the NNP, but pronounced
for the density-based estimators such as the \pofzw.  In
Sec.~\ref{sec:varcal}, we look at the impact of sample variance in calibration
of the photo-z error distributions, finding that it dominates shot-noise for
the scenarios we simulate.  Finally, in Sec. \ref{sec:disc} we examine the
dependence of our results on our choices of parametrizations.

\subsection{Spectroscopic redshift variance vs. photo-z variance}\label{sec:varvar}

\begin{figure}
\includegraphics[scale=0.3,angle=0]{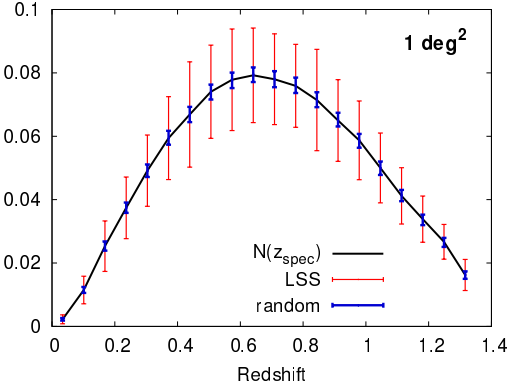}
\caption{Normalized spectroscopic redshift distribution for the full data. The
  red (light gray) error bars show the $1$-$\sigma$ variability in the
  redshift distribution for contiguous $1 {\textrm{ deg}}^2$ angular patches.
  The blue (dark gray) error bars show the variability in the redshift
  distribution assuming random samples of with the same mean number of objects
  as the $1 {\textrm{ deg}}^2$ patches.  We assume that only a 25\% random
  subsample of each patch is targeted for spectroscopy, yielding about
  $1.2\times 10^4$ galaxies per patch on average.  }
\label{fig:nzspec}
\end{figure}

Large-scale structure not only correlates the spatial distribution of
galaxies, but also correlates the distribution of galaxy types, colors, and
other properties.  For example, if there is a big galaxy cluster in some patch
on the sky, red galaxies will be over-represented in that patch.  Since red
galaxies typically have better photo-z's than blue galaxies, an estimate of
the redshift error distribution using this patch may not be representative of
the error distribution of the full sample.  In addition, objects in this
region will have a smaller dispersion in the quality of their redshifts than
predicted by Poisson statistics.  Because this extra systematic is indirectly
caused by the existence of large-scale structures, we refer to it as sample
variance of the photo-zs, to differentiate it from sample variance purely in
galaxy positions, which we hereafter refer to as the sample variance in the
spec-zs.

We use the conditional probabilities $\pzp$ and $\pzs$ to disentangle the two
sources of sample variance.  The key point is that $\pzs$ is sensitive to
changes in the $\zspec$ distribution, but not in the $\zphot$ distribution.
Conversely, $\pzp$ is only sensitive to changes in the $\zphot$ distribution,
but not in $\zspec$ (one can be convinced of this point by constructing
simple toy examples). 

We now estimate the variability of the error distributions across patches 
of the sky. For $\pzp$ we define the standard deviation about the mean
\begin{eqnarray}
\sigma(\pzp)=
\sqrt{\frac{ \sum_{\rm patches}  \left ( \pzp-\overline{\pzp}\right )^2}{N_{\rm patches}}} 
\label{eq:sigma_patches}
\end{eqnarray}
where $\overline{\pzp}$ is the mean 'leakage' (between the patches) of
galaxies from the spectroscopic bin centered at $z_s$ being registered as
having the photometric redshifts in the bin centered at $z_p$.  
We also introduce the equivalently defined quantity $\sigma(\pzs)$.  
We are interested in the increase in variability relative to the case of a 
random subsample, where effects of clustering due to the LSS have 
been zeroed out.

\begin{figure}
\centering
  \includegraphics[scale=0.35,angle=-90]{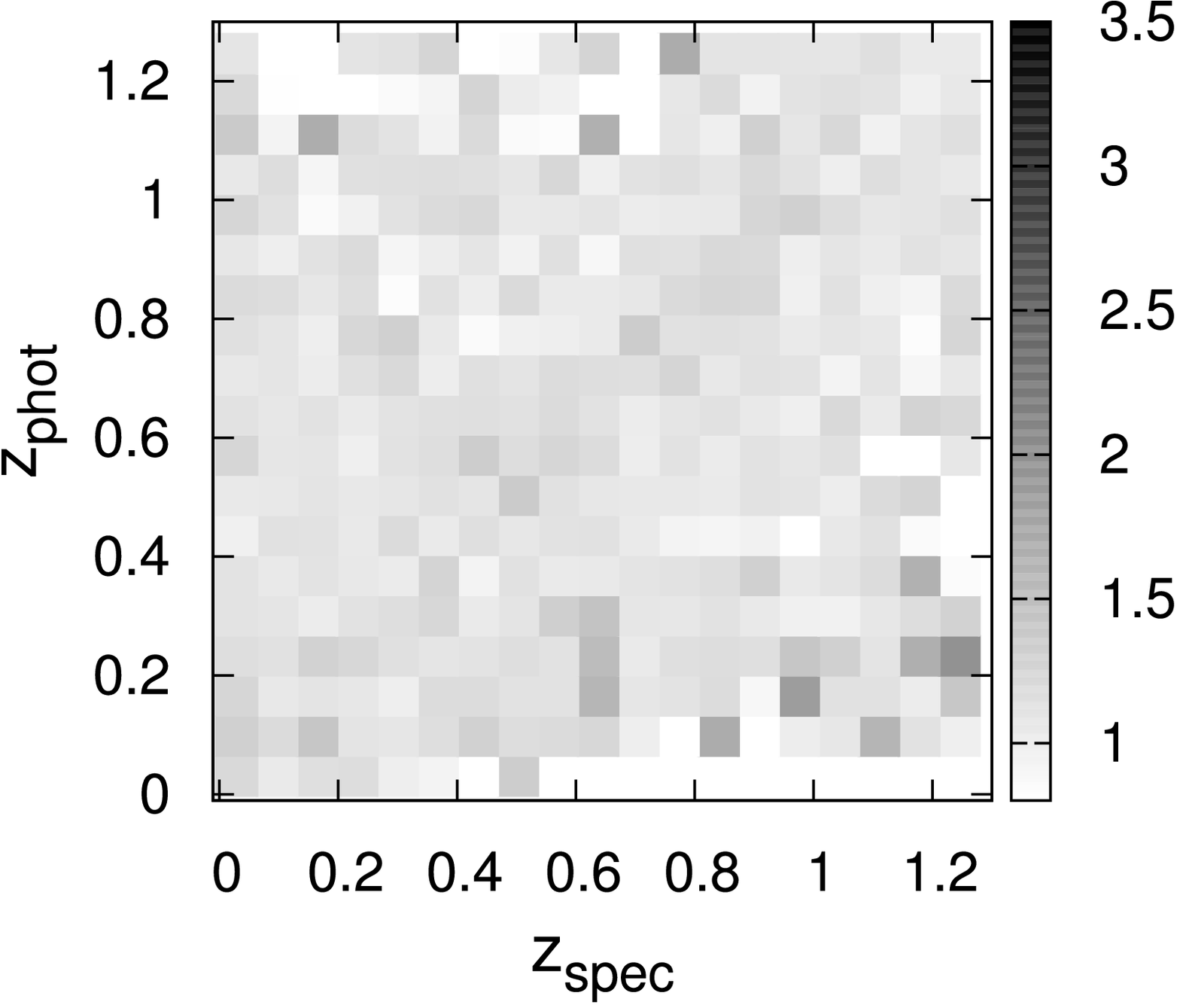}\vspace{0.2cm}
  \includegraphics[scale=0.35,angle=-90]{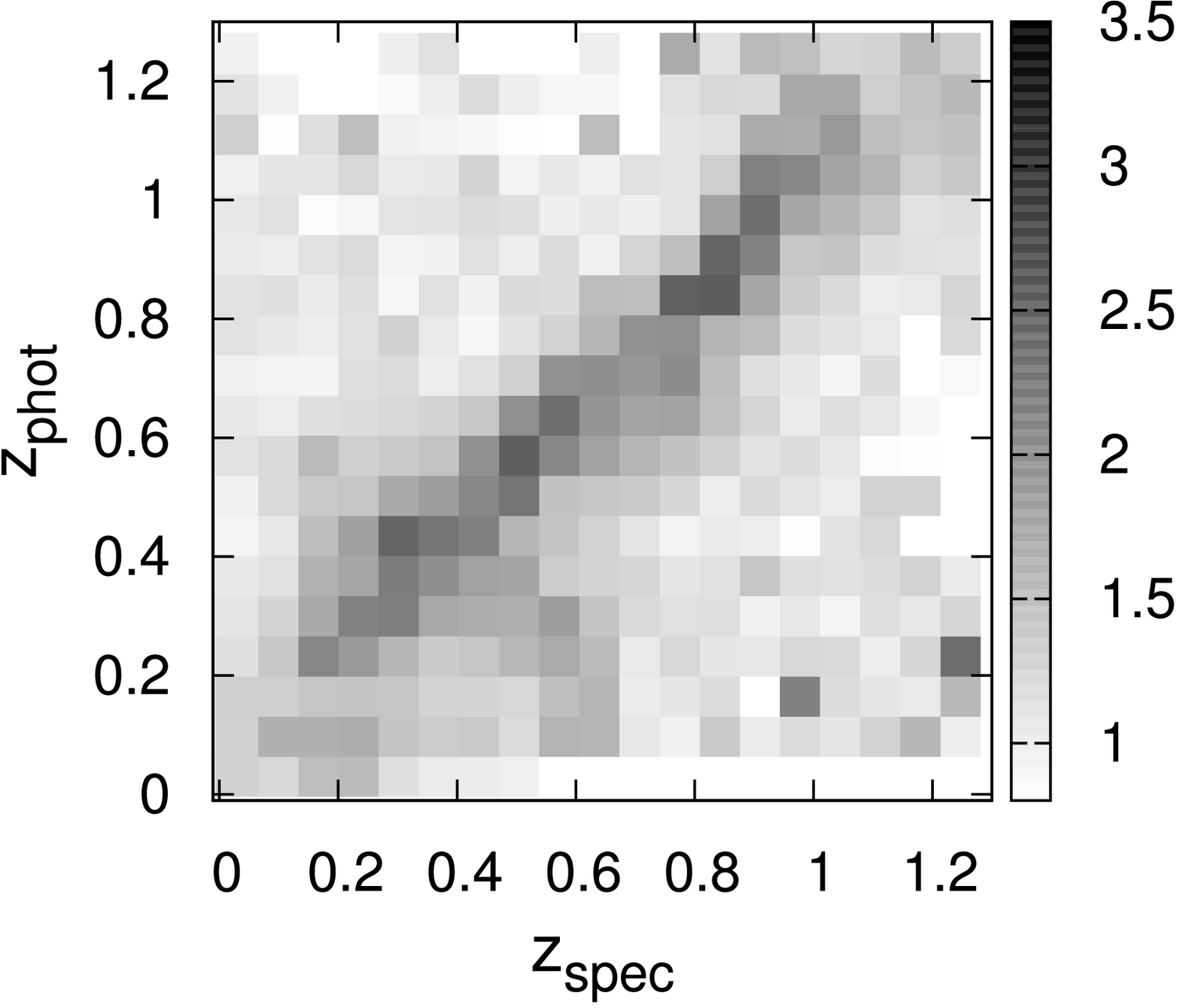}
\caption{ Top panel: Ratio of $\sigma(\pzp)$ (see
  Eq.~(\ref{eq:sigma_patches})) calculated for the $0.25 \degs$ LSS patches
  and the corresponding $0.25\degs$ random patches using template
  photo-z's. Bottom panel: same, but for $\sigma(\pzs)$.  The ratios are much
  bigger on the bottom plot than on the top, indicating that sample variance
  affects the spectroscopic redshifts much more than the photometric
  redshifts.  }
\label{fig:sigpratio}
\end{figure}

In the top panel of Fig.~\ref{fig:sigpratio} we show the {\it ratio} of
$\sigma(\pzp)$ calculated for the $0.25 \degs$ LSS patches and the
corresponding $0.25 \degs$ random-equivalent patches. In the bottom panel of
the same figure, we show the corresponding ratio for $\sigma(\pzs)$.  
We perform this test using the template photo-zs so as to isolate the importance
of sample variance on the calibration of the error matrices.  Comparing the
two plots, we see that sample variance of the photo-z's does not increase
appreciably between the random and the LSS patches, i.e. the ratios in each 
pixel are very close to unity.  
The sample variance of the spec-zs, on the other hand, shows
marked increase, as was already apparent from Fig.~\ref{fig:nzspec}.
In Sec. \ref{sec:improvnow} we show that the insensitivity of $\pzp$ to LSS 
can be used to reduce spectroscopic follow-up requirements.

\subsection{Sample variance in photo-z training}\label{sec:vartra}

\begin{table}
\begin{center}
\begin{tabular}{cc|cc|cc}
\hline\hline        \multicolumn{6}{c}{\rule[-2mm]{0mm}{6mm} Photo-z scatter and training set size }\\
\hline\hline \multicolumn{2}{c}{\rule[-2mm]{0mm}{6mm} }   &\multicolumn{2}{c}{\rule[-2mm]{0mm}{6mm} LSS} &\multicolumn{2}{c}{\rule[-2mm]{0mm}{6mm} Random}\\
\hline\hline   Area & Mean $N_{\rm gals}$ & \rule[-2mm]{0mm}{6mm} $\sigma_{poly}$ & \rule[-2mm]{0mm}{6mm} $\sigma_{p(z)}$ 
& \rule[-2mm]{0mm}{6mm} $\sigma_{poly}$    
& \rule[-2mm]{0mm}{6mm}  $\sigma_{p(z)}$ \\\hline\hline
\rule[-2mm]{0mm}{6mm}$6 \degs$   &$7.4 \times 10^4$  &0.099  &0.104 & 0.099  &0.104  \\
\rule[-2mm]{0mm}{6mm}$1 \degs$   &$1.2 \times 10^4$  &0.106  &0.129 & 0.105  &0.129  \\
\rule[-2mm]{0mm}{6mm}$0.25 \degs$&$3.0 \times 10^3$  &0.114  &0.162 & 0.113  &0.163  \\\hline\hline
\end{tabular}
\caption{$1$-$\sigma$ scatter of the polynomial photo-zs (averaged over all
  training iterations) and mean $1$-$\sigma$ width of the \pofzw s, (averaged
  over all training iterations).  These mean scatters are shown for different
  patch areas and training set sizes.  For comparison, the mean scatter of the
  template-fitting photo-zs is 0.157. Note that the LSS does not affect the
  photometric redshift statistics significantly, but the total number of
  galaxies in the training set does.  }
\label{tab:ztrain}
\end{center}
\end{table}

In this section we examine the effects of sample variance in the training of
photo-zs. We find that the commonly reported scatter in the photo-z estimation
is affected by the shot noise but not by sample variance.

Table \ref{tab:ztrain} shows the average photo-z scatter of the photometric
sample for the polynomial method as well as the average width of the \pofzw s.
The photo-z scatter is defined as the standard deviation (around zero) of the
$P(\zphot-\zspec)$ distribution.  The average mean width of the \pofzw\ is
defined as the average, over all training iterations, of the mean $1$-$\sigma$
width of the \pofzw s of the galaxies in the photometric sample.  Comparison
of the corresponding 'LSS' and 'Random' columns in the Table shows that
large-scale structure does not affect the photo-z or \pofzw\ statistics
significantly.  The training set size is important, however, as larger
training sets have lower shot noise.  For the polynomial photo-z's, we see a
$12\%$ degradation in the scatter between the $6 \degs$ and $0.25 \degs$
cases.  The \pofzw s are much more sensitive, with a degradation of $63\%$.

\begin{figure*}
\includegraphics[scale=0.23,angle=-90]{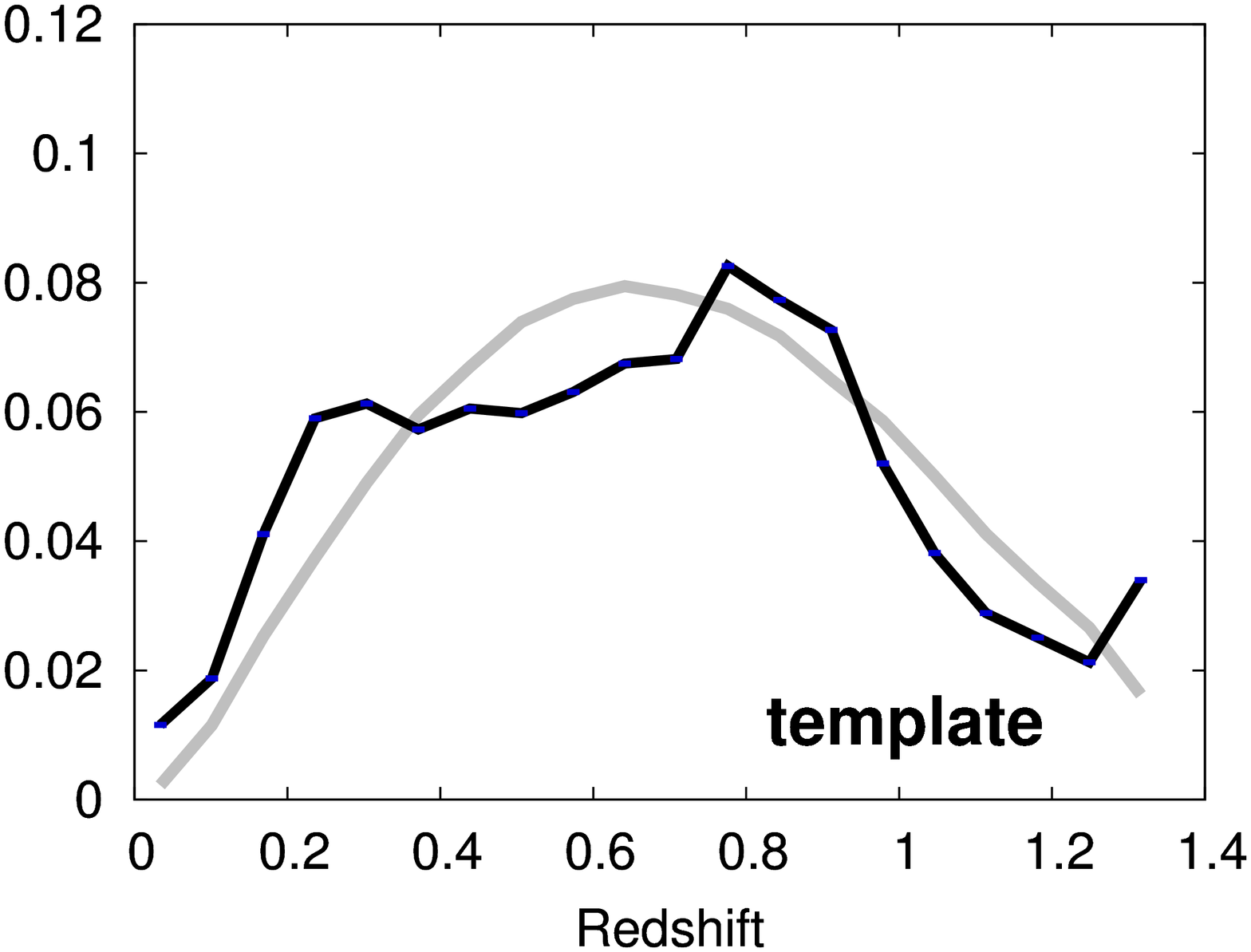}\hspace{0.cm}
\includegraphics[scale=0.23,angle=-90]{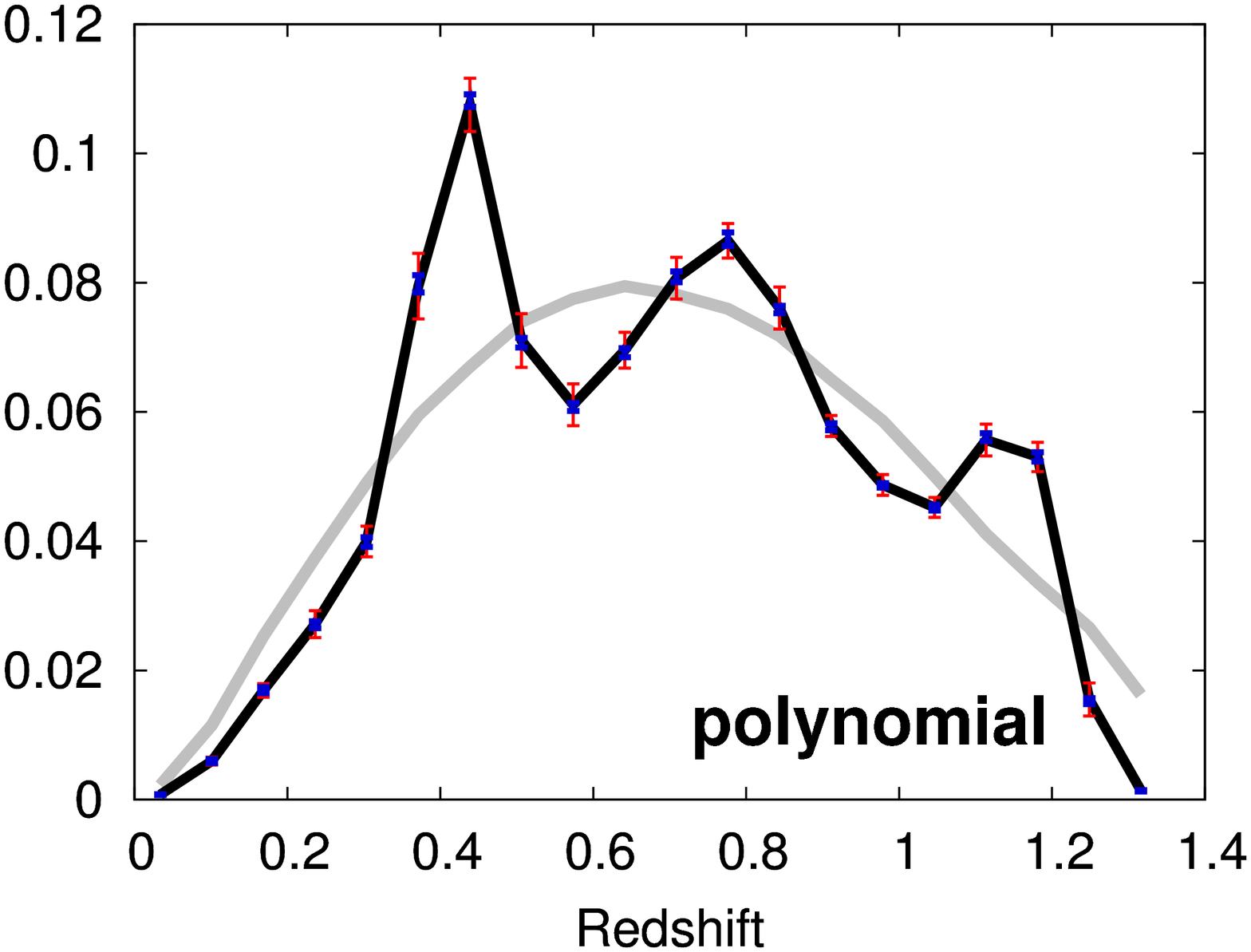}\hspace{0.cm}
\includegraphics[scale=0.23,angle=-90]{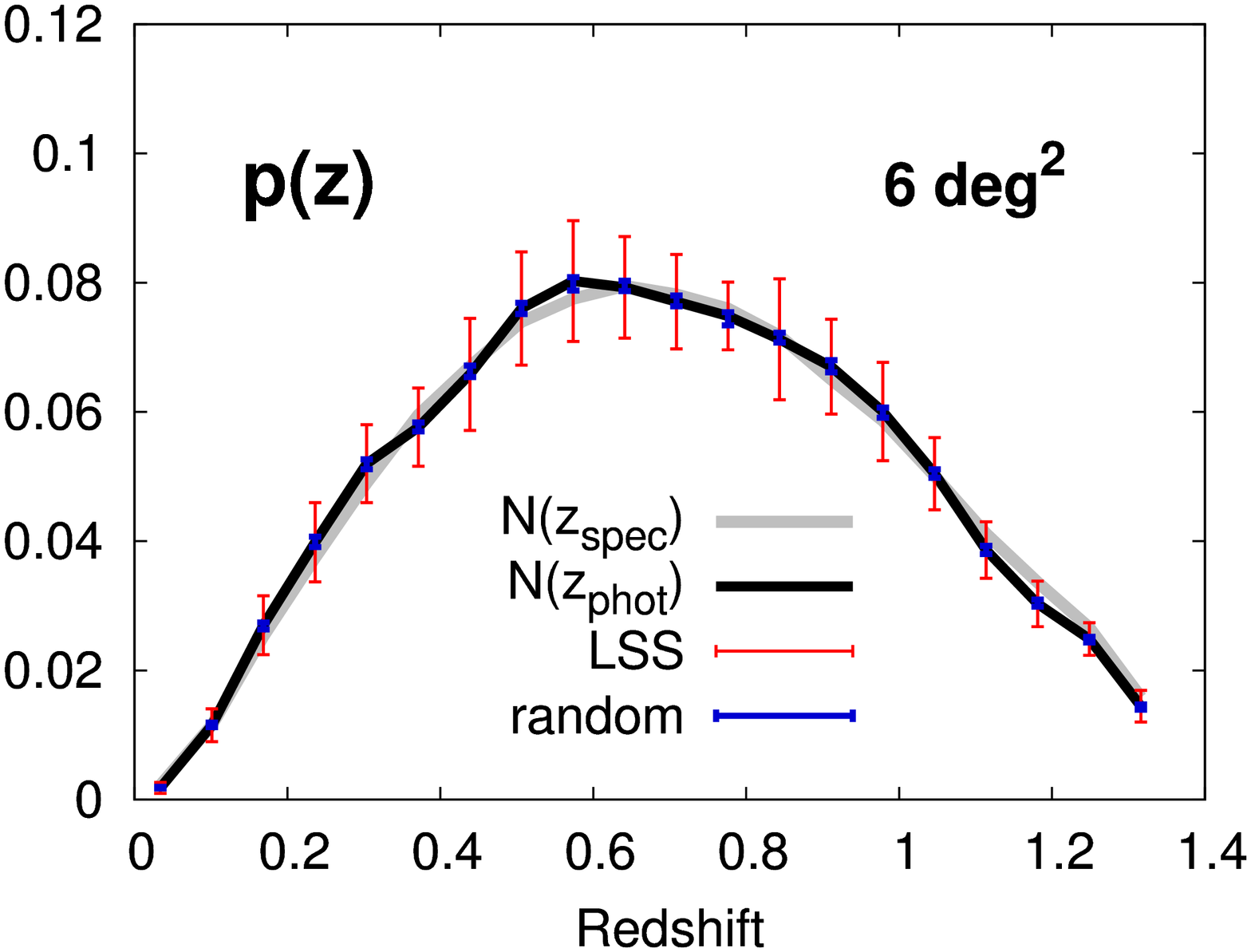}
\caption{Redshift distribution estimates using the (left) template-fitting,
  (center) polynomial and (right) \pofzw\ estimator.  The true redshift
  distribution is shown in gray, and the estimates are in black.  The weights
  estimate is not shown as it is indistinguishable from the true redshift
  distribution.  The red (light gray) error bars 
  shows the
  $1$-$\sigma$ variability of the estimates for the $6 \degs$ patches.  
  The hardly visible blue (dark gray) error bars show the corresponding 
  error bars derived using the random equivalent subsamples. 
  Note that the template fitting and polynomial methods produce very 
  precise but highly biased estimates of the redshift distribution.  }
\label{fig:nzp}
\end{figure*}

This demonstrates that one can significantly decrease the variance of the
recovered redshifts by fitting the redshift-observable relation
(e.g.\ using the polynomial method) instead of using a pure
density estimator (e.g.\ the \pofzw) -- however, this
comes at the cost of {\it biasing} the recovered redshift distribution, as
seen in Fig.~\ref{fig:nzp}. What are the options, then, for improving
 the latter class of methods?
To reduce the width of the \pofzw\ one can either use
repeat observations to decrease the mean neighbor separation in the training
set, decrease the number of nearest-neighbors used, or adopt a fit to the
redshift-observable density distribution in the neighborhood of each galaxy.
We leave these explorations for a future work.

The message of this section is that the intrinsic uncertainty of photo-zs is
much greater than any systematic introduced by large-scale structure, so that
there is no significant degradation of photo-z scatter {\it itself} by
  using training sets obtained from pencil beam surveys.  However, the
commonly reported photo-z scatter is not sufficient to gauge biases on
cosmological parameters.  Below we will show that sample variance introduced
by the LSS does in fact lead to significant biases in cosmological parameter
estimates.

\begin{figure*}
\includegraphics[scale=0.23,angle=-90]{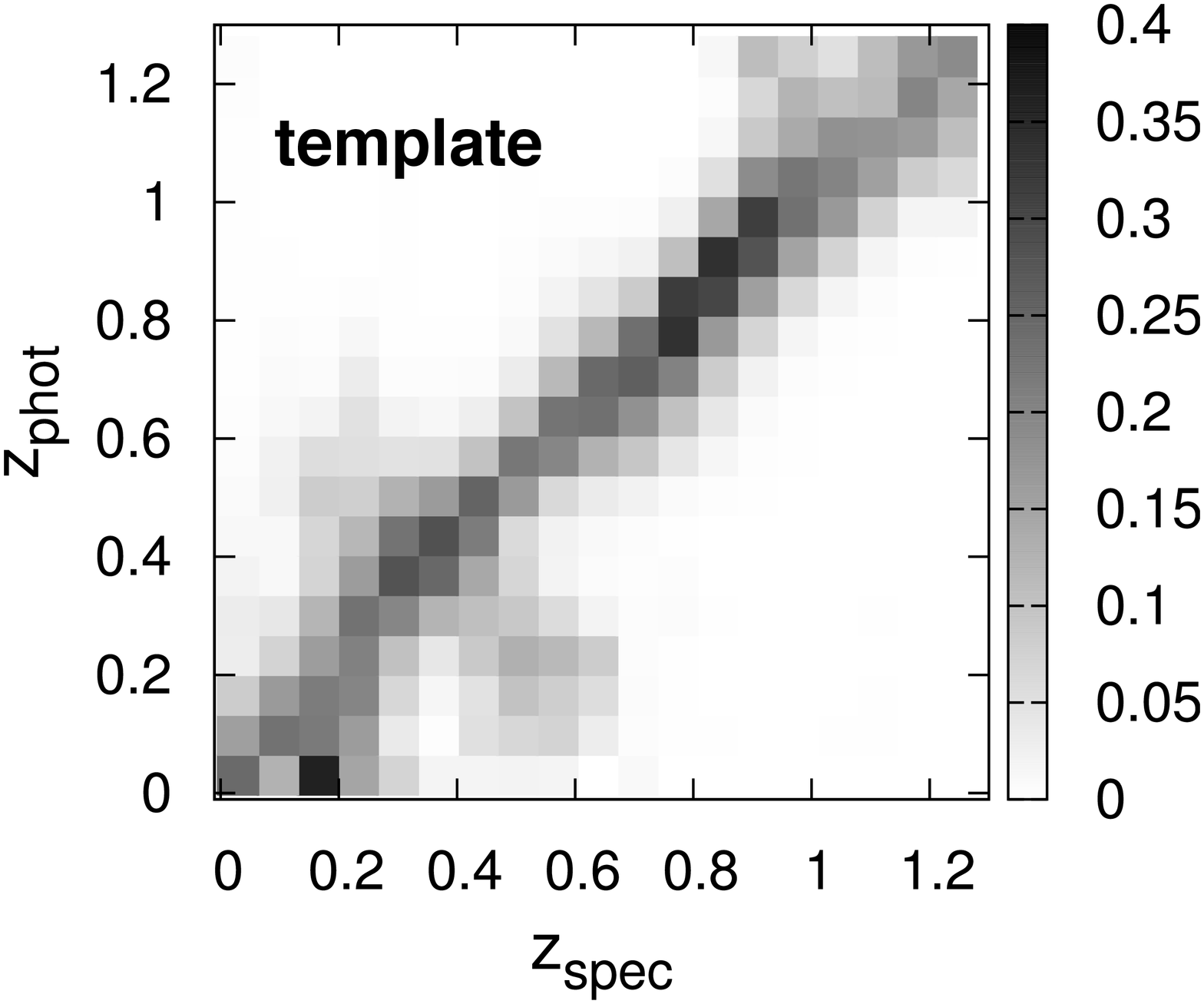}\hspace{0.3cm}
\includegraphics[scale=0.23,angle=-90]{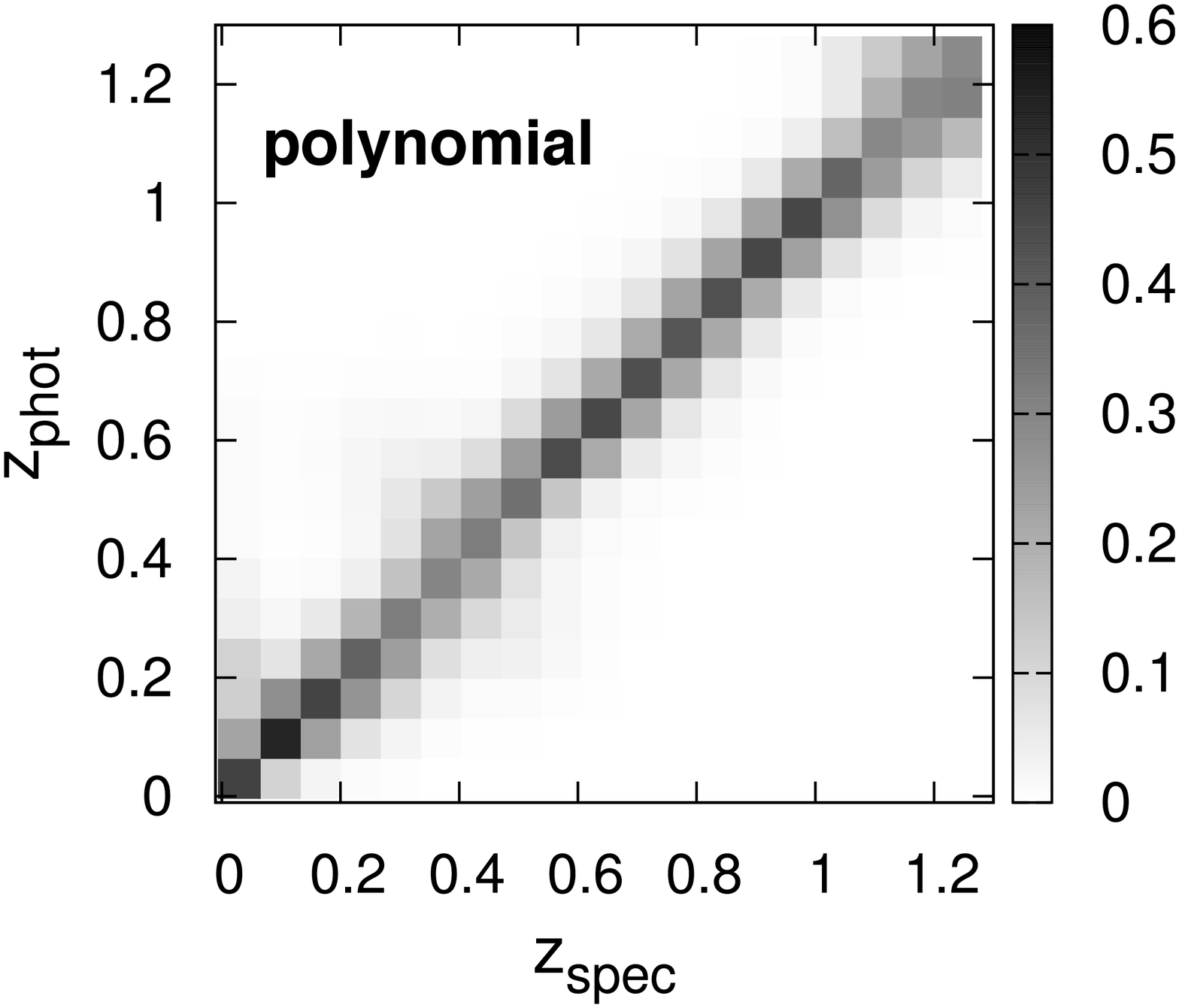}\hspace{0.3cm}
\includegraphics[scale=0.23,angle=-90]{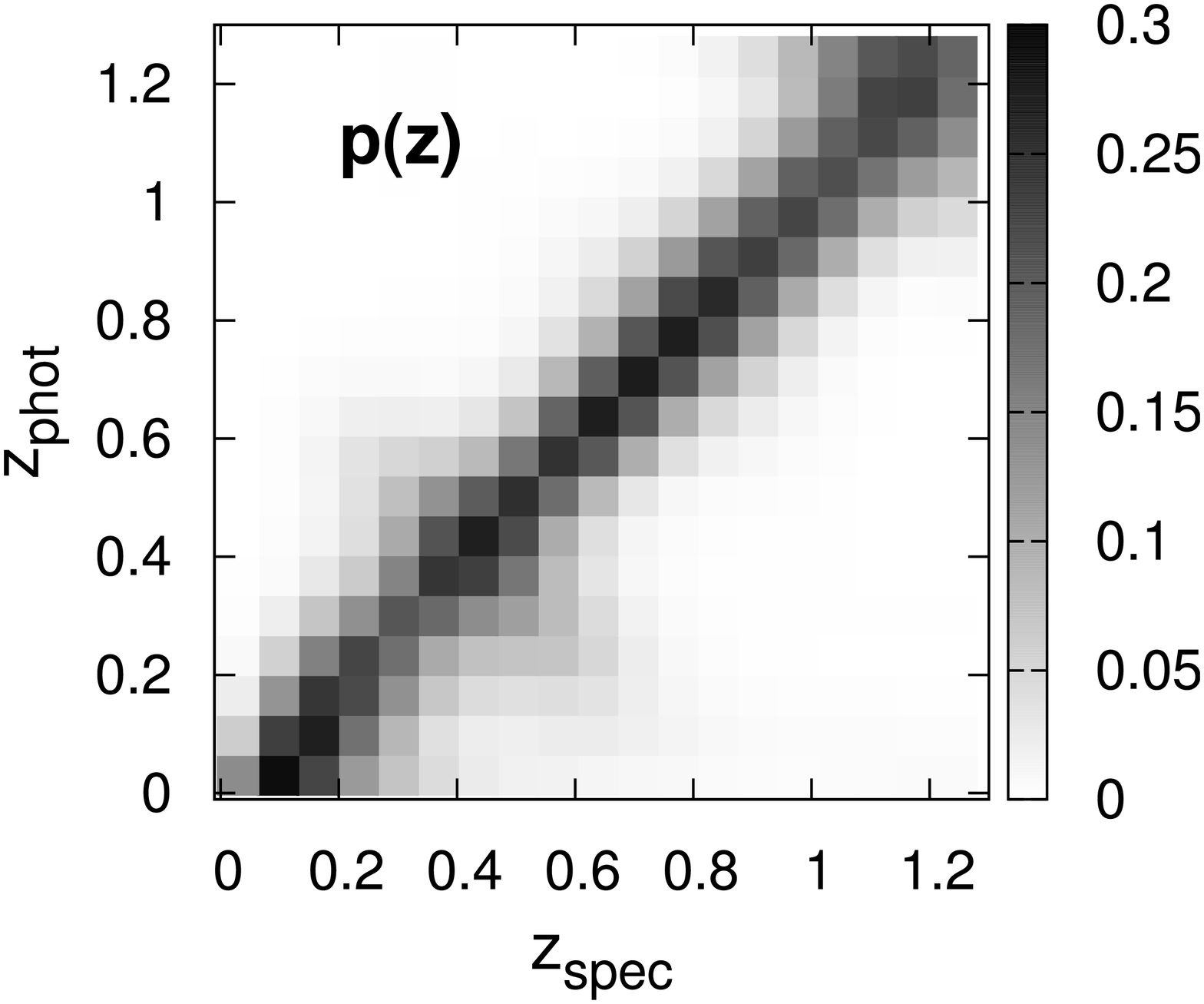}
\caption{Mean $\pzs$ for the three methods. The template is on the left,
  polynomial at the center, and \pofzw\ on the right.  For the polynomial and
  \pofzw, the mean $\pzs$ depend on the training size.  We show the $6 \degs$
  result for both. Note the different scales in the three plots.  }
\label{fig:pzpzs}
\end{figure*}

\subsection{Sample variance in photo-z calibration}\label{sec:varcal}

In this Section, we describe how the sample variance in the spectroscopic
parameters biases the calibration of the photo-z error distributions 
(i.e. the $\pzs$), and how this translates into bias in cosmological 
parameters.
The main metric we use to quantify the cosmological bias is the 
fractional bias in the equation of state $w$.
We define the fractional bias as the absolute
bias in $w$ obtained from Eq.~(\ref{bias1}) divided by the fiducial statistical
error
\begin{equation}
{\delta w\over\sigma(w)}
\end{equation}
where the marginalized statistical error in the equation of state is, recall,
$\sigma(w)= 0.035$ for the DES+Planck combination (see Sec.~\ref{sec:wl}).

We begin by examining a single patch in Sec.~\ref{sec:p37} and then discuss
statistics of the biases for all the calibration patches in the simulation.

\subsubsection{Case study: Patch 37}\label{sec:p37}

To understand how fluctuations in the redshift distribution of the calibration
sample affect the estimation of $\pzs$ and the resulting cosmological
biases, we focus on a single $1 \degs$ calibration patch, Patch 37 (out of,
recall, 225 total patches).  We choose this patch (which happens to be 37th in
our ordering) randomly, but check that it is  fairly typical, with total
fractional bias well within the $1$-$\sigma$ limits of the fractional bias
distribution for the two methods we investigate.

Before we get to the details we review a result (covered in BH10) which we
will utilize. Fig.~\ref{fig:nz37} shows the ratio of biases
in the dark energy equation of state $w$ divided by its statistical error
induced by each individual photo-z error corresponding to a fixed
contamination $\pzs$ of 0.01 in each{$(\zspec$,$\zphot)$ bin.  The points to
note are that cosmological biases generally worsen with distance from the
$\zphot=\zspec$ line, i.e.\ as the photo-z error becomes 'more
catastrophic'. Conversely, contamination is relatively harmless at low
$\zphot$ or at $\zphot$ near the survey median.

Now we are ready to examine Patch 37.  The examination consists of two steps.
In step 1, we look into how the differences between the overall redshift
distribution and the redshift distribution of Patch 37 affect the estimation
of the error distribution $\pzs$ for the polynomial and template methods.  In
step 2, we look at how the errors in the estimation of $\pzs$ in any
given $(z_s, z_p)$ bin propagate to biases in the dark energy equation of state
$w$.
\begin{itemize}

\item \textbf{Step 1: Patch 37 redshift biases}. 
  Fig.~\ref{fig:nz38} shows the spectroscopic redshift distribution
  of the whole survey (i.e.\ of the photometric sample) $\nzsphot$ in black
  color, as well as that of Patch 37, $\nzpts$, in blue (gray).  The deviations of
  the redshift distribution of Patch 37 from that of the full survey directly
  affect the estimation of $\pzs$, regardless of photo-z method.  The
  top-row panels of Fig.~\ref{fig:p37} show the difference of $\pzs$ for
  the full sample and Patch 37 (the calibration sample) for the polynomial
  method (top left) and template method (top right).  
  Comparing Fig.~\ref{fig:nz38} to the top-row panels of
  Fig.~\ref{fig:p37}, we see that each downward fluctuation of $\nzpts$
  relative to $\nzsphot$ translates into a negative $\delpzs$ for the
  corresponding $\zspec$ column regardless of photo-z method used. The converse
  is also true: if $\nzpts$ overestimates $\nzsphot$ at a given $\zspec$ bin,
  then $\delpzs$ will be biased high in that $\zspec$ column as well.  

\item \textbf{Step 2: Patch 37 biases in $\mathbf{w}$}. The bottom-row panels
  of Fig.~\ref{fig:p37} show the corresponding fractional biases in the dark
  energy equation of state $w$ in each ($\zspec$,$\zphot$) bin.  For each
  $(z_s, z_p)$ bin, the fractional bias in $w$ is essentially a product
  between the sensitivity in fractional $w$ bias to unit redshift errors
  (shown in Fig.~\ref{fig:nz37}) and the actual redshift
  bias (shown in the left-column panels of Fig.~\ref{fig:p37} for the two
  photo-z methods). Even though the sensitivities for fixed contamination are
  smallest near the $z_s \approx z_p$ diagonal, the actual values of $\delpzs$
  are largest near the diagonal.  Overall, the latter effect  wins, as the
  right panels of Fig.~\ref{fig:p37} show, and the biases in $w$ are
  contributed largely -- though not exclusively -- by $\delpzs$ errors near
  the diagonal, $z_s\approx z_p$.  A noticeable exception is the bin near
  $\zspec=0.4$, $\zphot=1.3$, in the polynomial results (left column).
  Overall, the contribution of this bin lowered the overall fractional bias in
  $w$, which turns out to be $\delta w/\sigma(w)=0.27$ for the polynomial
  method and 0.52 for the template method.  Hence, if it wasn't for the big
  negative bias in that bin, the polynomial would have lost to the template
  method in this patch!  The conclusion is that the final $w$ bias is the
  result of several cancellations, which reduce the importance of the choice
  of photo-z method.  However, it is desirable that photo-zs be accurate
  because it implies that the $\pzs$ will more diagonal, which, for comparably
  stable methods, implies smaller biases in $w$.  And perhaps most
  importantly, better photo-z's imply better fiducial constraints, which our
  analysis is not sensitive to.

\end{itemize}

\begin{figure*}
\begin{minipage}[b]{0.4\linewidth}
    \includegraphics[scale=0.4,angle=-90]{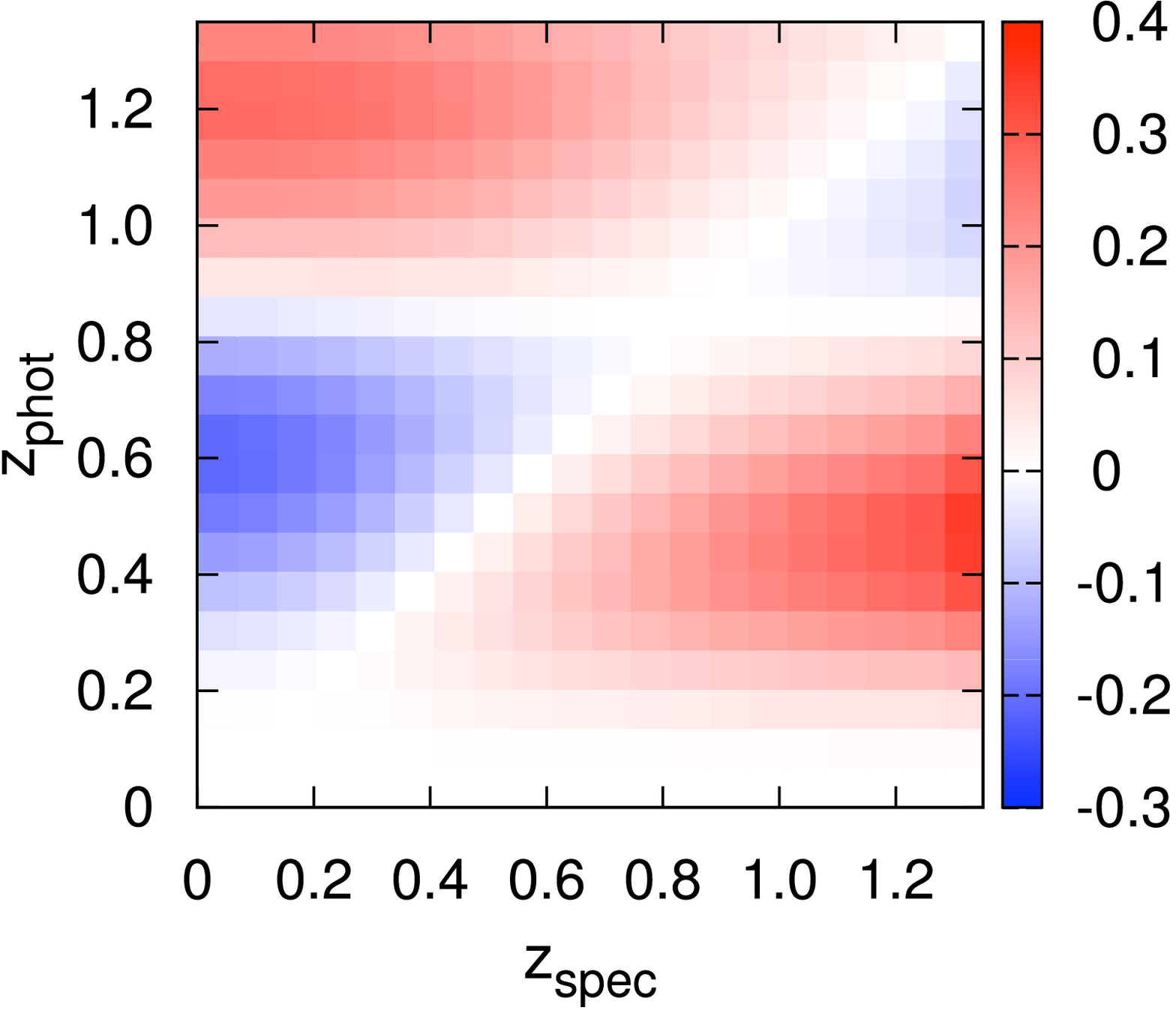}
\caption{Bias/error ratio in the dark energy equation of state,
       $\delta w/\sigma(w)$, for a fixed contamination of 0.01 as a function
       of position in $\zphot -\zspec$ space.
     }
\label{fig:nz37}
\end{minipage}
\hspace{0.5cm}
\begin{minipage}[b]{0.4\linewidth}
\includegraphics[scale=0.3,angle=-90]{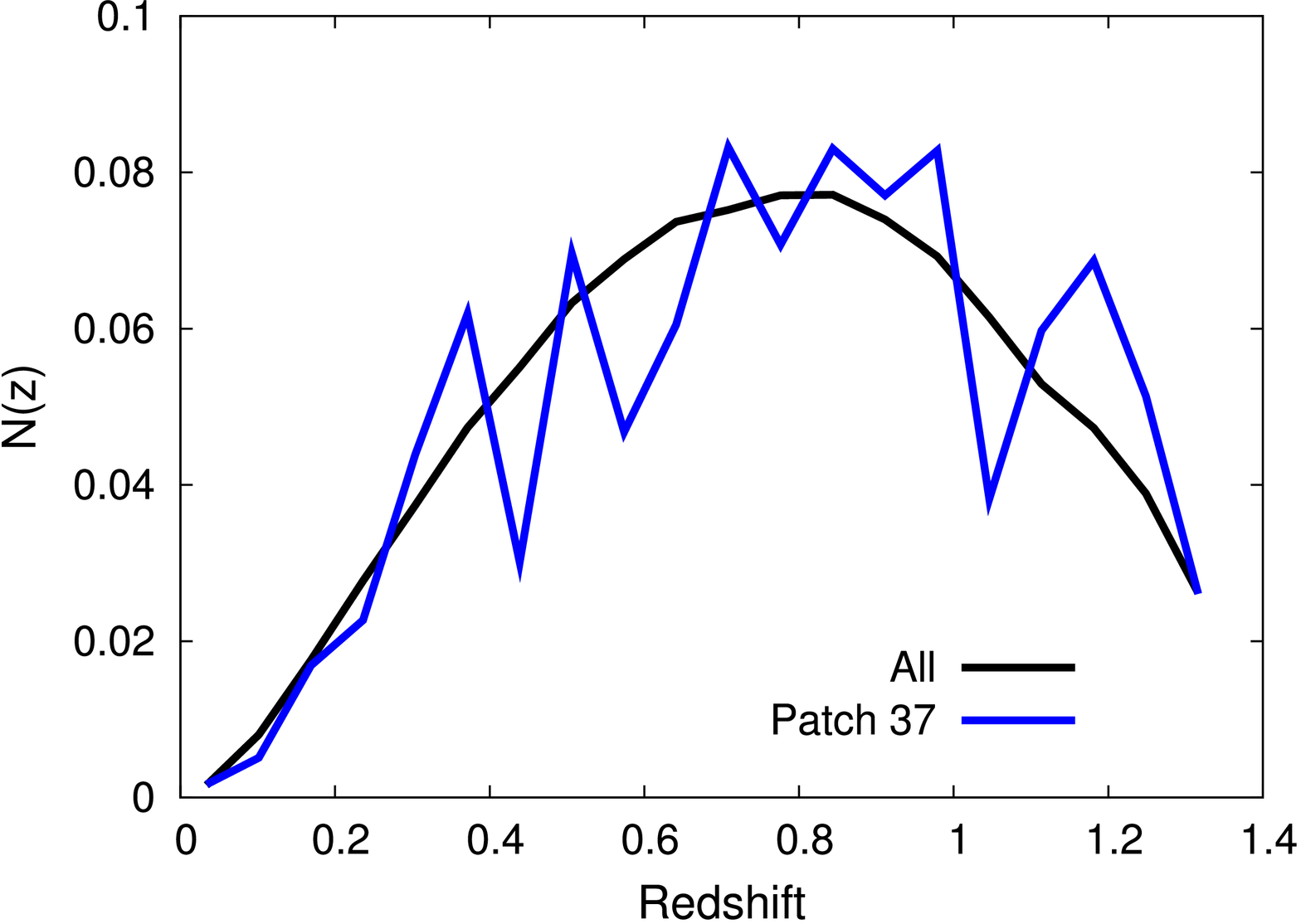}
\caption{ spectroscopic redshift distribution of the 
       whole survey (i.e. the photometric sample), $\nzsphot$, in  black, 
       and of Patch 37, $\nzpts$, shown in  blue.}
\label{fig:nz38}
\end{minipage}
\end{figure*}

\begin{figure*}
\begin{minipage}[t]{80mm}
\includegraphics[scale=0.32,angle=-90]{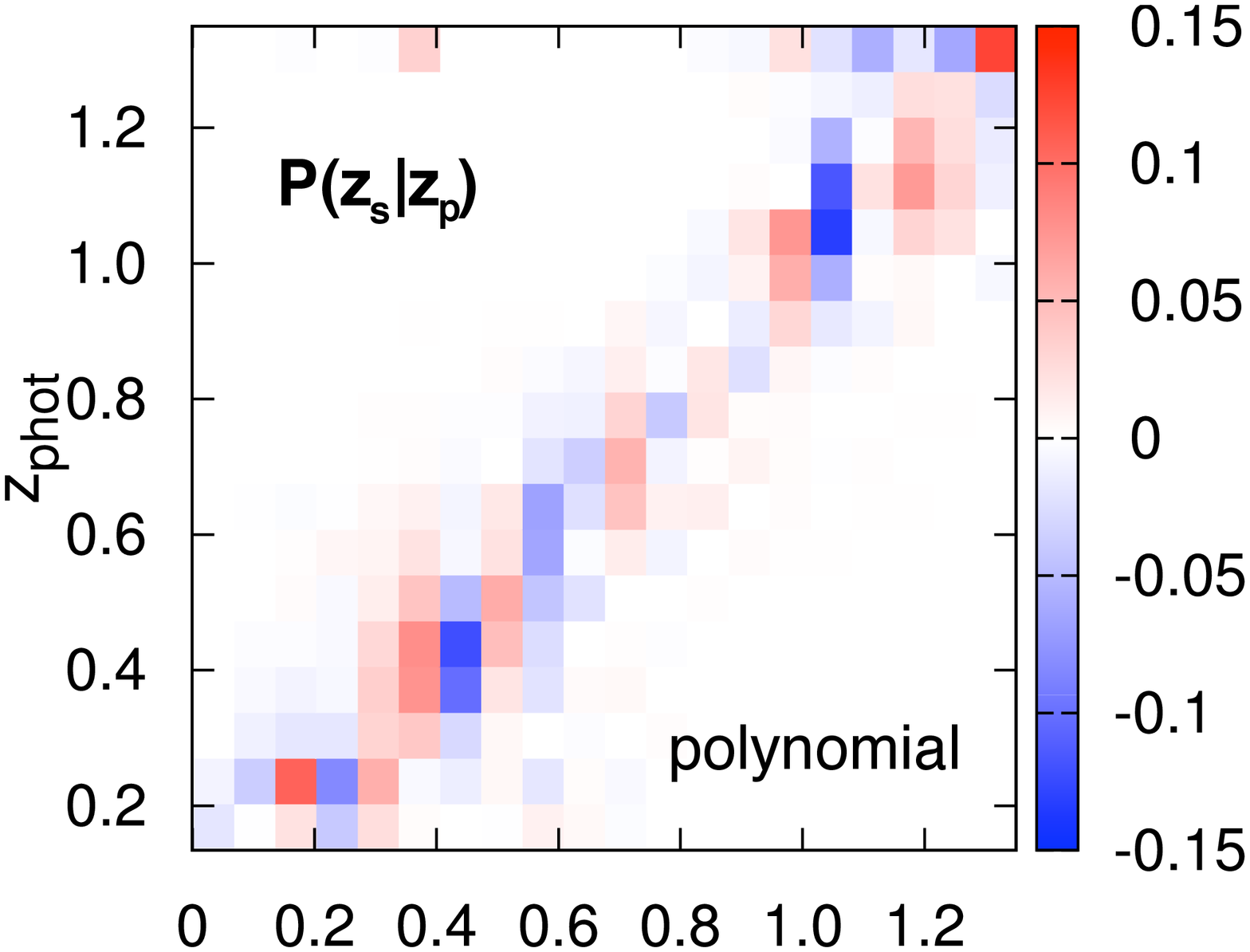}\vspace{0.8cm}
\includegraphics[scale=0.4,angle=0]{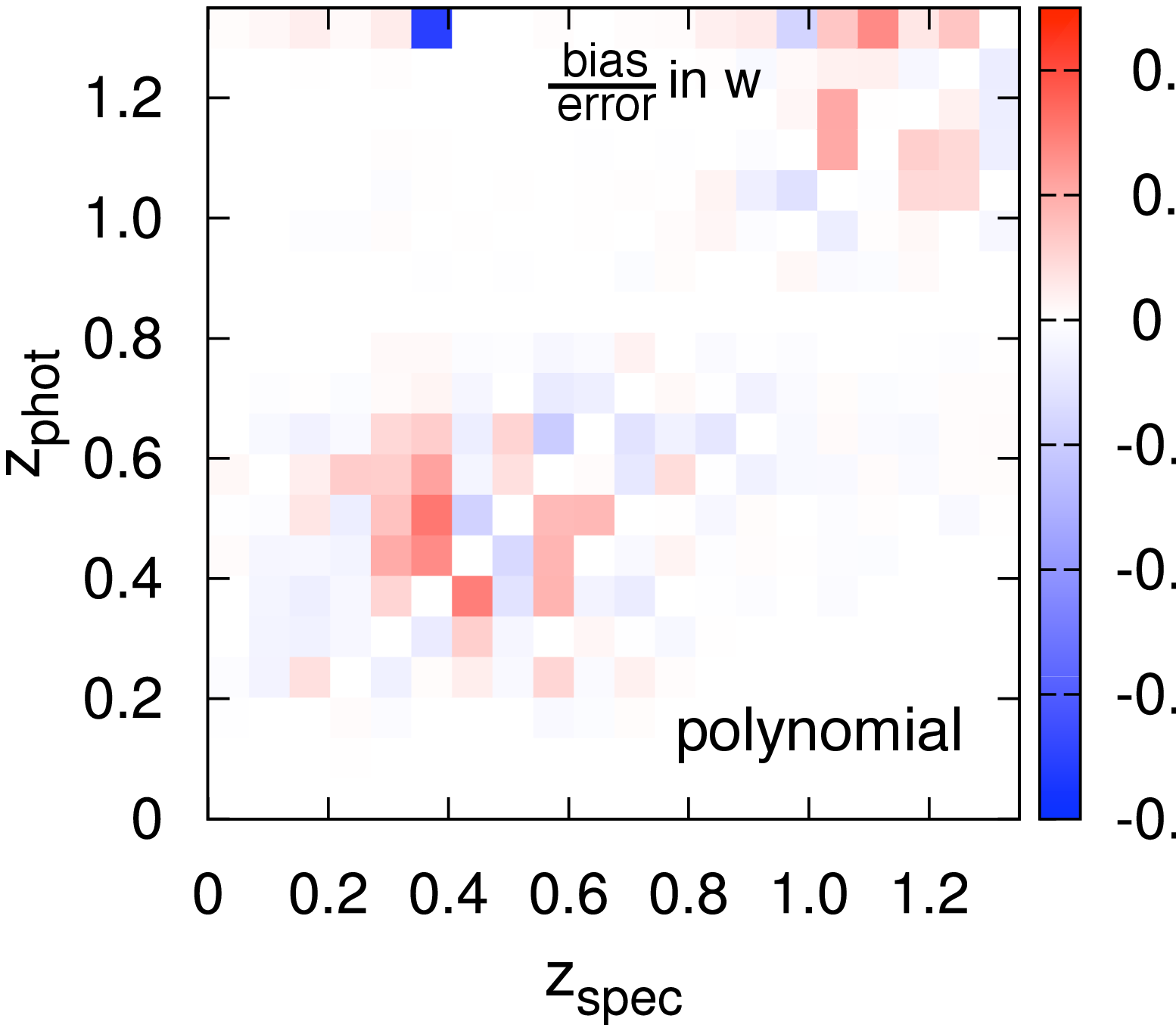}
 \end{minipage}
\begin{minipage}[t]{80mm}
\includegraphics[scale=0.32,angle=-90]{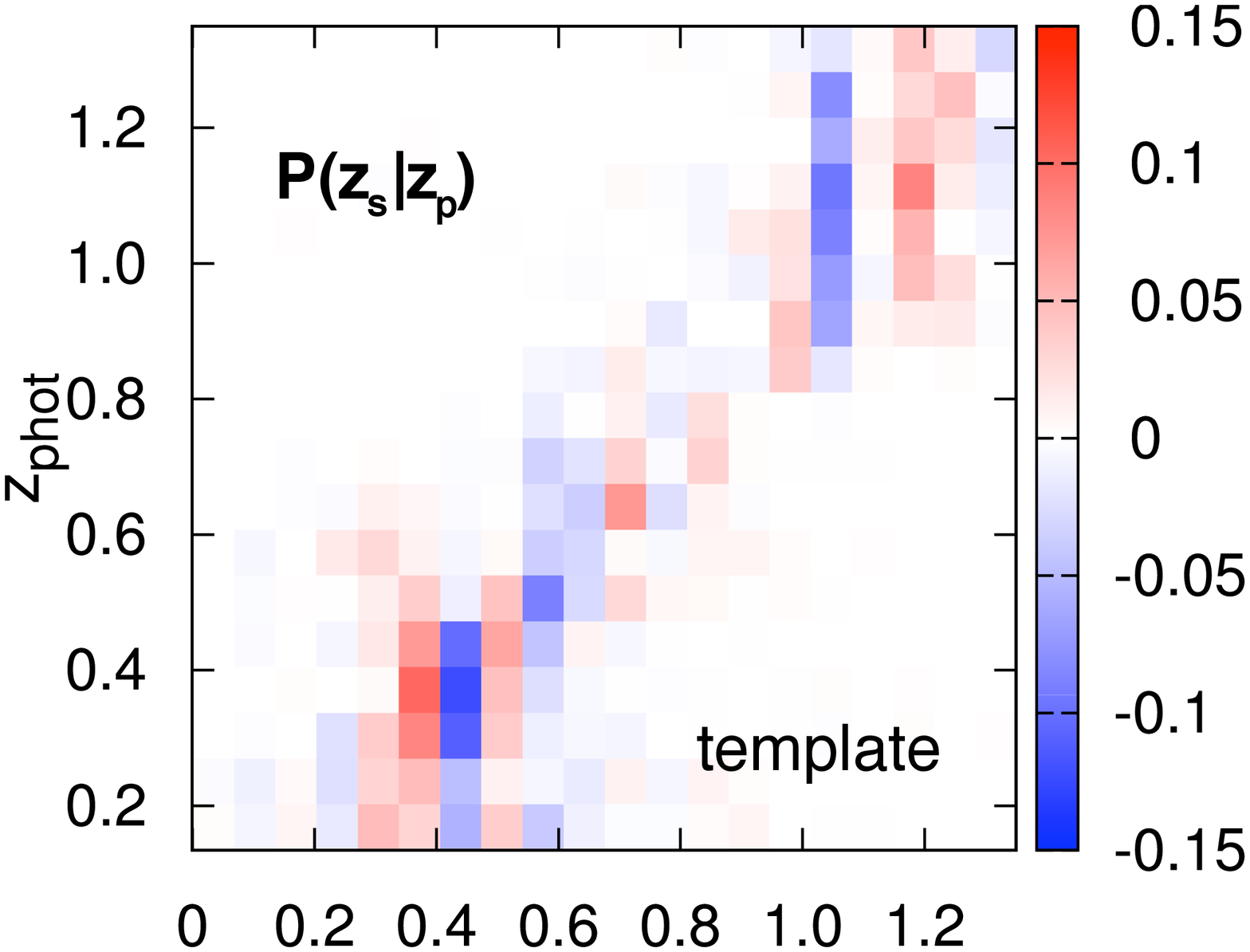}\vspace{0.8cm}
\includegraphics[scale=0.4,angle=0]{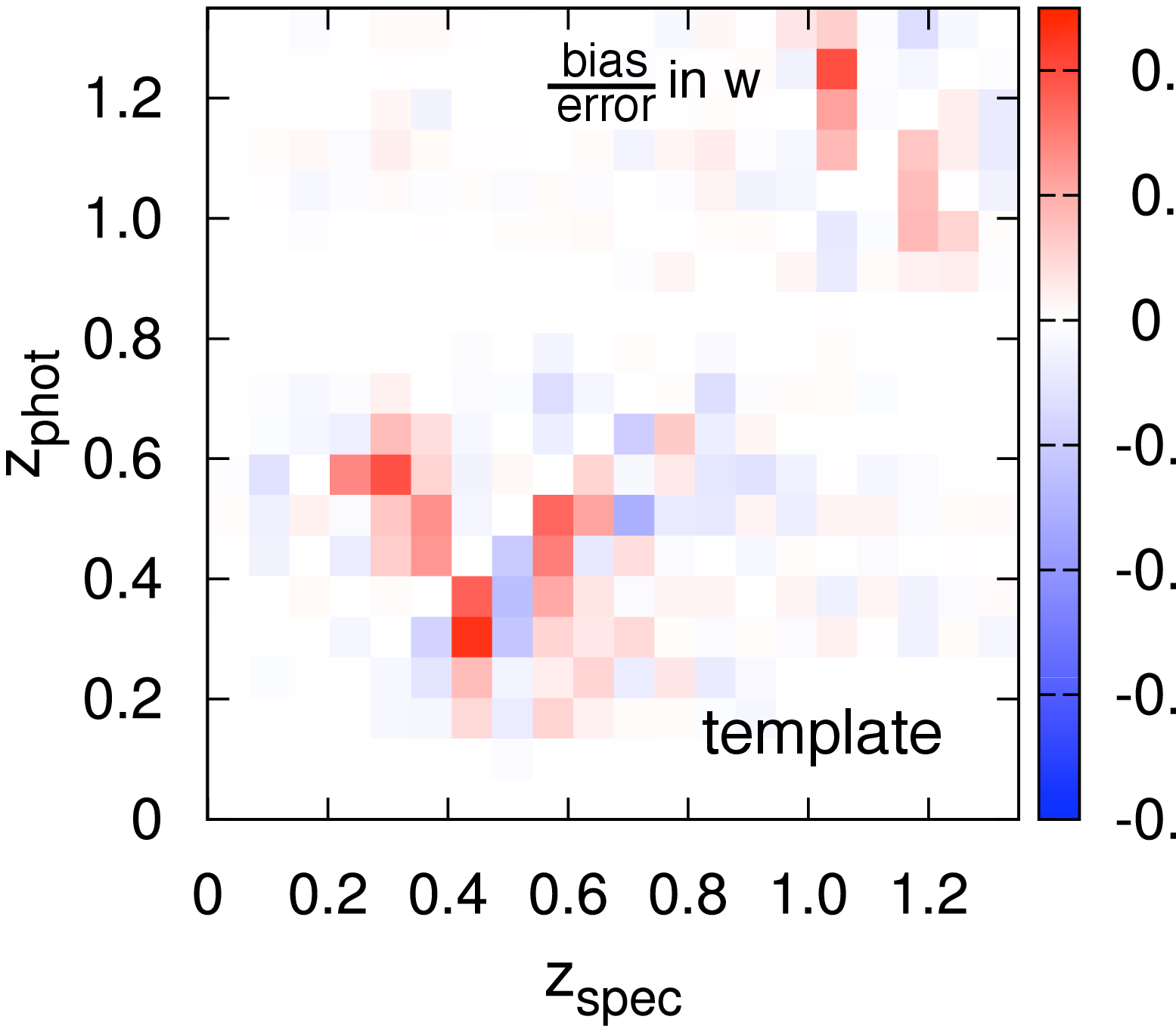}
 \end{minipage}
\caption{Biases in Patch 37. The top-row panels shows the difference of $\pzs$
  for the photometric and calibration samples for the polynomial (top left panel) and
  template (top right panel) method.  The bottom-row panels show the corresponding
  contribution to bias/error ratio in the dark energy equation of state $w$ due to
  photometric redshift errors in each {$\zspec$, $\zphot$} bin. 
  The fractional biases in $w$ shown in the bottom row panels are equal to the 
  product of the photometric redshifts errors (shown in the top row panels) and 
  the sensitivity to a fixed photometric redshift (shown in Fig.~\ref{fig:nz37}).}
\label{fig:p37}
\end{figure*}

\subsubsection{Statistics of the biases in $w$}\label{sec:wstats}

In this Section we examine statistics of the biases in $w$ when different
patches are used for training and/or calibration of the photo-zs.
Fig.~\ref{fig:biasbias} shows the distribution of the fractional biases when
using the \pofzw\ and template-fitting estimators as a function of the biases
obtained when the polynomial technique is used.  The top panel shows the $1
\degs$ LSS case, and the bottom plot shows the $1 \degs$ random equivalent.
Clearly, biases in $w$ introduced by sample variance for the different methods
are very correlated while those introduced by Poisson fluctuations alone are
not.  This suggests that one cannot reduce the effects of sample variance by
simply combining estimates based on different photo-z methods.

In Table \ref{tab:wbias}, we show the mean fractional bias in the equation of
state $w$, its $\sigma_{68}$ statistics, and the median total shift in
chi-squared (defined below) corresponding to the full-dimensional cosmological
parameter space.  We define $\sigma_{68}$ as the range encompassing 68\% of
the area of the distribution of $|\delta w|/\sigma(w)$, where $\delta w$ is
the bias in the equation of state in any given patch and $\sigma(w)$ is the
marginalized statistical error in the equation of state. Moreover, we define
the total chi-square as

\begin{equation}
\chisq = (\delta {\bf p})^T F\, {\bf p}
\label{eq:chisq}
\end{equation}
where $\delta {\bf p}$ is a six-dimensional vector containing cosmological
parameter biases and $F$ is the (statistical-only) Fisher matrix defined in
Eq.~(\ref{eq:Fisher}).  We then define $\chisqm$ to be the median of the
distribution of $\chisq$.

We find that the distribution of fractional biases are typically reasonably
Gaussian, in the sense that our definition of $\sigma_{68}$ matches the
standard deviation of the fractional bias distribution (without the absolute
value) to a few percent, and an equivalent definition of $\sigma_{95}$ is
quite close to twice the standard deviation.  In Sec.~\ref{sec:guide}, we will
assume the distribution of fractional biases is Gaussian to estimate follow-up
requirements for the DES survey.

Actual spectroscopic calibration samples should be comprised of several sets
of patches of sky.  Ideally, the patches should be separated enough so as to
be statistically independent.  Because of the small size of our simulation it
is not possible to combine many independent patches; recall, our simulation
covers only $\sim 15\deg$ on a side.  As a simple alternative, we combine
several randomly selected patches to create the spectroscopic training and
calibration sample.  We consider two scenarios, one comprised of patches 120
of $1/8 \degs$ with each galaxy selected with probability of 0.03125 - with
average total of $2.4\times 10^4$ galaxies.  The other scenario is comprised
of 180 patches of $1/32 \degs$ with galaxies selected with probability 0.125,
and with the average total of $3.4 \times 10^4$ galaxies.  We repeat the
procedure for generating these combined samples several times to generate the
statistics shown in Table \ref{tab:wbiascomb}.

The point we want to make is that, in the more realistic scenarios with
calibration samples coming from separate patches, all of the photo-z methods
we tested yield very similar results.  Combining patches randomly is far from
ideal, hence the bias statistics presented in Table \ref{tab:wbiascomb} are
pessimistic.  We consider the spectroscopic requirements with optimal patch
selection in Sec.~\ref{sec:guide}.

The conclusions of this Section are:

\begin{itemize} 

\item The LSS and random-equivalent cases lead to very different bias statistics. 
  Conversely, differences between the photo-z methods do not affect the bias
  statistics considerably.  In particular, when many patches are combined, the
  photo-z estimators perform nearly identically.

\item The \pofzw\ method is the most sensitive to sample variance.  This is
  expected because it is a purely density-based estimator, and it degrades the
  fastest as the area and size of the training set decrease.  However,
  comparing the statistics of the \pofzw\ for different areas in the random
  equivalent cases suggests that the \pofzw\ estimator is not as sensitive to
  shot noise.  Moreover, the \pofzw\ method is the only method that yields a
  perfect reconstruction of the overall redshift distribution in the limit of
  large area of spectroscopic samples.

\item The polynomial-fitting method appears to have slightly larger mean fractional bias
  than the \pofzw\ and  template-fitting in the cases shown in 
  Table \ref{tab:wbias}.
  However, the mean fractional bias is significantly smaller than the
  $\sigma_{68}$ width in all cases.  In addition, the polynomial technique
  outperforms the other methods in almost all scenarios, suggesting that 
  use of a training set yields improvements superior to any bias introduced 
  by using the same patch to train and calibrate the photo-zs.
  We believe that the conclusion that one can use the same sample to train
  and calibrate photo-zs should hold for other training-set-dependent photo-z
  techniques provided the method has some control for the degrees of freedom
  it utilizes and thereby avoid biases due to over-fitting.

\end{itemize}

\begin{figure}
\includegraphics[scale=0.33,angle=-90]{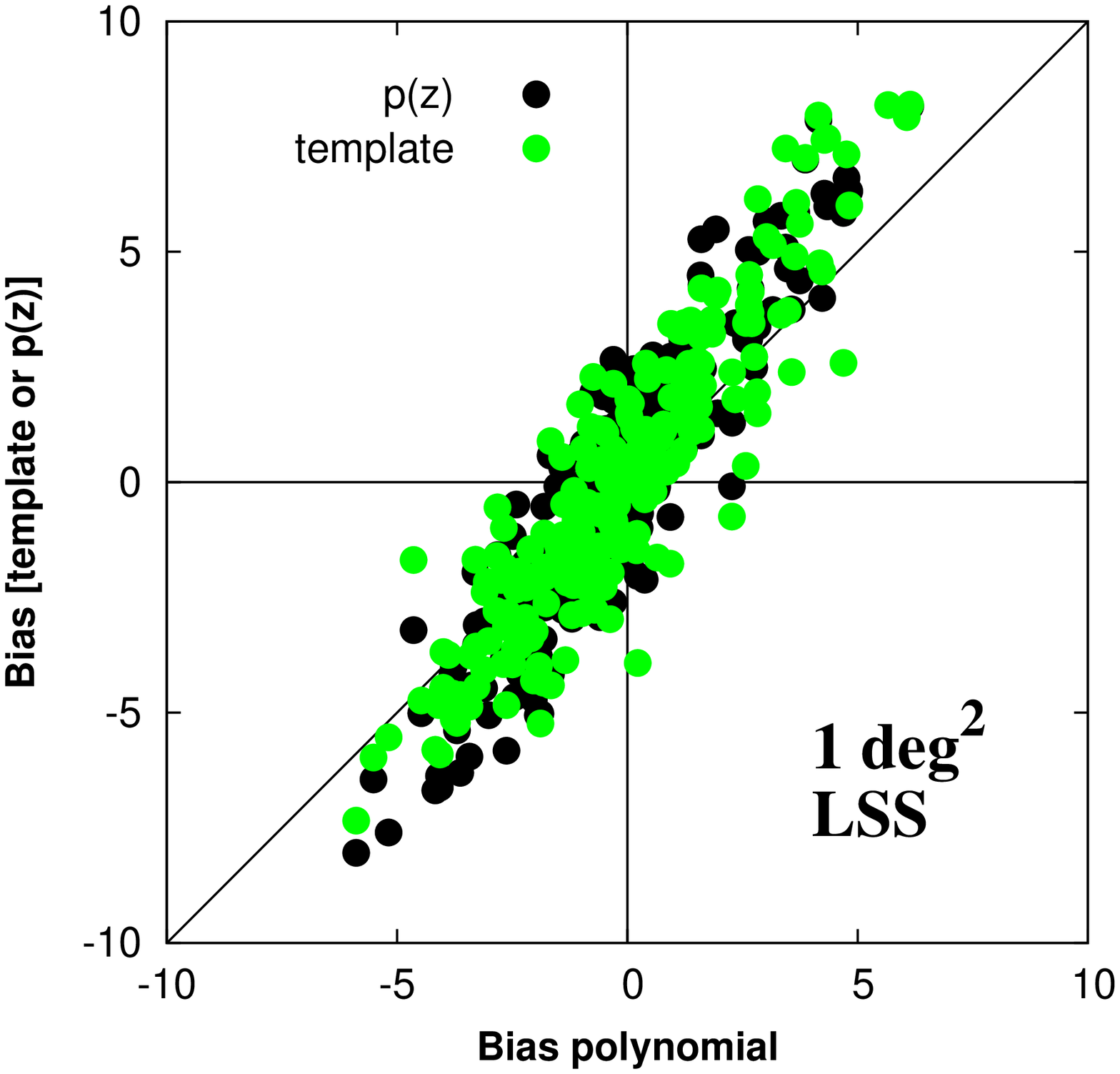}\vspace{1.cm}
\includegraphics[scale=0.33,angle=-90]{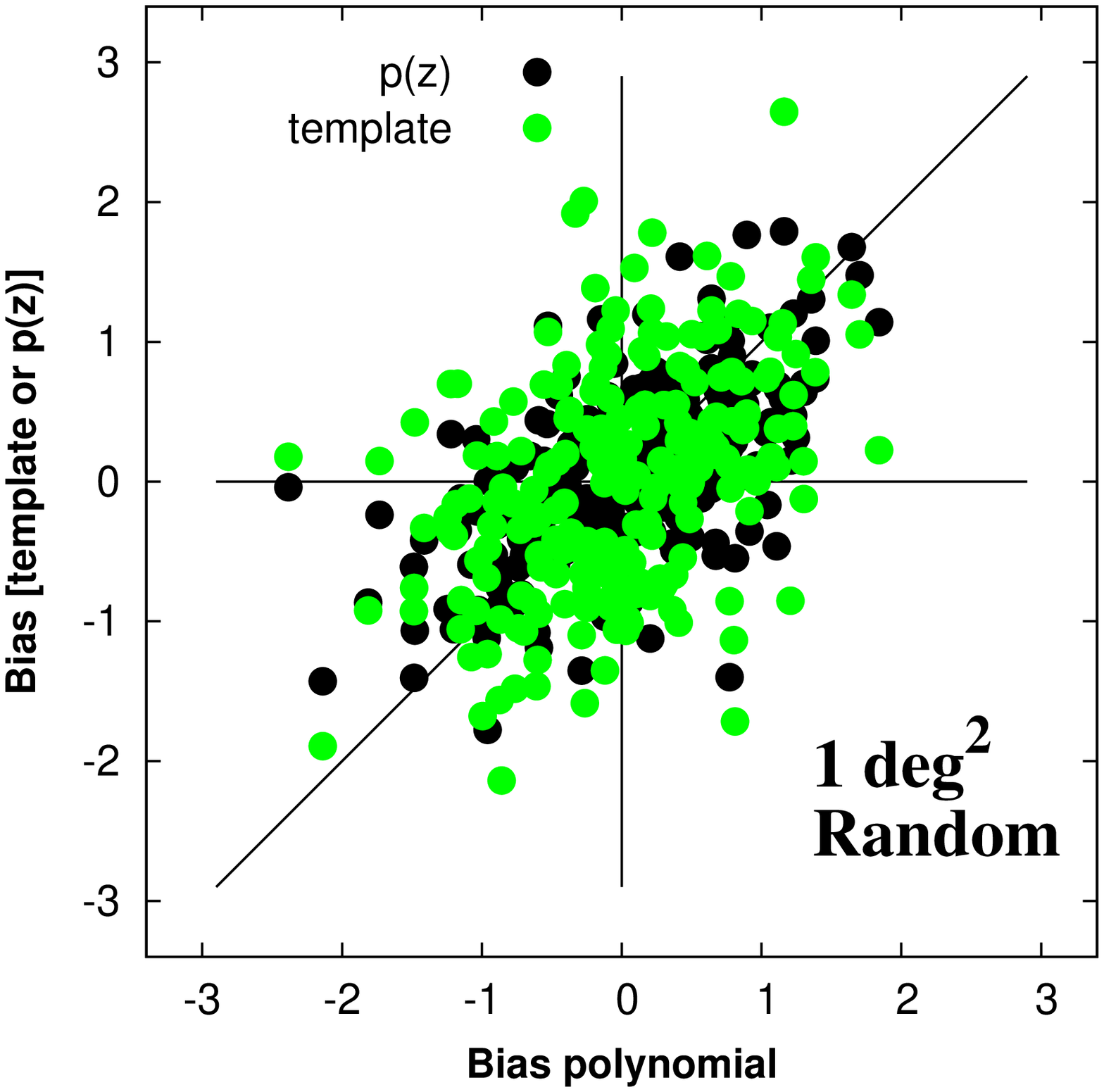}\vspace{1.0cm}
\caption{Fractional biases in $w$ (i.e.\ the
  bias/error ratios in $w$) for the different $1 \degs$ patches used to train
  and/or calibrate the photometric redshifts. The top panel shows that errors
  in different photo-z methods produce correlated biases in the equation of
  state $w$ in the presence of the LSS. The x-axis indicates the fractional
  bias in $w$ for the polynomial estimator, while the y-axis shows the
  corresponding bias for template estimator (black points) and the
  \pofzw\ estimator (green points).  The bottom panel shows the random
  equivalent patches where the correlation is much less pronounced. 
}
\label{fig:biasbias}
\end{figure}

\begin{table}
\begin{center}
\begin{tabular}{c|ccc|ccc}
\hline\hline        \multicolumn{7}{c}{\rule[-2mm]{-3mm}{6mm} \bf{Bias in w} }\\
\hline\hline  $\mathbf{6 \degsb}$  &\multicolumn{3}{c}{\rule[-2mm]{-3mm}{6mm} LSS} 
                                   &\multicolumn{3}{c}{\rule[-2mm]{-3mm}{6mm} Random}\\
\hline          Technique 
& \rule[-2mm]{-3mm}{6mm} \dwmean%
& \rule[-2mm]{-3mm}{6mm} $\sigma_{68}$ 
& \rule[-2mm]{-3mm}{6mm} $\chisqm$ 
& \rule[-2mm]{-3mm}{6mm} \dwmean%
& \rule[-2mm]{-3mm}{6mm} $\sigma_{68}$  
& \rule[-2mm]{-3mm}{6mm} $\chisqm$ 
\\\hline
\rule[-2mm]{-3mm}{6mm}Template    & 0.04 & 2.56 & 3.14   & 0.04 & 0.44 & 0.14 \\
\rule[-2mm]{-3mm}{6mm}Polynomial  & -0.07 & 1.53 & 2.04   & -0.04 & 0.39 & 0.12 \\
\rule[-2mm]{-3mm}{6mm}\pofzw        & 0.05 & 2.33 & 2.56 &0.07 & 0.31 & 0.10 \\
\hline  $\mathbf{1 \degsb}$  &\multicolumn{3}{c}{\rule[-2mm]{-3mm}{6mm} \textcolor{white}{$a$.addddaaaaaaa} } &\multicolumn{3}{c}{\rule[-2mm]{-3mm}{6mm} \textcolor{white}{aaaddddaaaaaa} }\\
\hline
\rule[-2mm]{-3mm}{6mm}Template    &-0.04 & 3.75 & 7.36  &0.01 & 0.92 & 0.75  \\
\rule[-2mm]{-3mm}{6mm}Polynomial  &-0.19 & 2.96 & 4.74 & 0.00 & 0.93 & 0.64 \\
\rule[-2mm]{-3mm}{6mm}\pofzw        &-0.01 & 3.99 & 9.05 &0.029 & 0.78 & 0.50  \\
\hline  $\mathbf{1/4 \degsb}$  &\multicolumn{3}{c}{\rule[-2mm]{-3mm}{6mm}\textcolor{white}{$a$.addddaaaaaaa} } &\multicolumn{3}{c}{\rule[-2mm]{-3mm}{6mm} \textcolor{white}{aaaddddaaaaaa} }\\
\hline
\rule[-2mm]{-3mm}{6mm}Template    &0.03 & 4.61 & 16.4 & -0.15 & 1.9 & 2.9\\
\rule[-2mm]{-3mm}{6mm}Polynomial  &-0.11 & 3.99 & 10.3 &-0.17 & 1.7 & 2.2 \\
\rule[-2mm]{-3mm}{6mm}\pofzw      &0.07 & 5.88 & 32.3  &-0.10 & 2.0 & 3.0\\\hline\hline
\end{tabular}
\caption{Mean fractional bias in $w$ (i.e.\ mean of $\delta w/\sigma(w)$)
  and $\sigma_{68}$ (i.e.\ width of the $|\delta w|/\sigma(w)$ distribution)
  for the different techniques, assuming patches of area 6, 1, $1/4 \degs$ for
  training and calibration or a random subsample with the same number of
  galaxies.  The $\chisqm$ column indicates the median value (among all
  patches) of $\chisq $ of the fit over all cosmological parameters; see
  Eq.~(\ref{eq:chisq}).  }
\label{tab:wbias}
\end{center}
\end{table}

\begin{table}
\begin{center}
\begin{tabular}{c|ccc}
\hline\hline   \multicolumn{4}{c}{\rule[-2mm]{0mm}{6mm} \bf{Bias in w (combined random patches) }}\\
\hline\hline   \multicolumn{4}{c}{\rule[-2mm]{0mm}{6mm}  $\mathbf{1/8 \degsb}$ - fraction = 0.03125}\\[-0.20cm]
  \multicolumn{4}{c}{\rule[-2mm]{0mm}{6mm}  120 patches - $\overline{N} =2.2 \times 10^4$}\\[-0.1cm]
\hline          Technique & \rule[-2mm]{0mm}{6mm} \dwmean& \rule[-2mm]{0mm}{6mm} $\sigma_{68}$ & \rule[-2mm]{0mm}{6mm} $\chisqm$   \\\hline
\rule[-2mm]{0mm}{6mm}Template    & 0.12 & 0.84 & 0.59   \\
\rule[-2mm]{0mm}{6mm}Polynomial  & -0.16 & 0.76 & 0.54   \\
\rule[-2mm]{0mm}{6mm}\pofzw        & 0.05 & 0.84 & 0.54   \\\hline
\multicolumn{4}{c}{\rule[-2mm]{0mm}{6mm}  $\mathbf{ 1/32 \degsb}$ - fraction = 0.125}\\[-0.20cm]
\multicolumn{4}{c}{\rule[-2mm]{0mm}{6mm}  180 patches - $\overline{N} =3.4\times 10^4 $}\\[-0.1cm]
\hline          Technique & \rule[-2mm]{0mm}{6mm} \dwmean& \rule[-2mm]{0mm}{6mm} $\sigma_{68}$ & \rule[-2mm]{0mm}{6mm} $\chisqm$  \\\hline
\rule[-2mm]{0mm}{6mm}Template    &0.19 & 0.76 & 0.41  \\
\rule[-2mm]{0mm}{6mm}Polynomial  &-0.10 & 0.62 & 0.29  \\
\rule[-2mm]{0mm}{6mm}\pofzw        &0.12 & 0.74 & 0.39  \\\hline\hline
\end{tabular}
\caption{ Mean fractional bias in $w$ (i.e.\ $\delta w/\sigma(w)$) and
  $\sigma_{68}$ (i.e.\ width of the $|\delta w|/\sigma(w)$ distribution) for the
  different techniques, assuming 120 randomly selected patches of area $1/8
  \degs$or 180 patches of area $1/32 \degs$ were used for training and
  calibration.  Galaxies selected from the $1/8 \degs$ and the $1/32 \degs$
  patches with probabilities 0.125 and 0.03125, respectively.  
  The $\chisqm$ column indicates the median
  $\chisq $ of the fit over all cosmological parameters.}
\label{tab:wbiascomb}
\end{center}
\end{table}

\subsection{Dependence on simulations and parametrizations}\label{sec:disc}

In this section we discuss some of our choices of survey parameters.

\subsubsection{Dependence on intrinsic ellipticity}\label{sec:disce}
For most of the results shown on this paper, we have assumed
  the optimistic value of $\langle\gamma_{\rm int}^2\rangle^{1/2}=0.16$ for
  the rms intrinsic ellipticity.  The effective intrinsic ellipticity is
  somewhat difficult to estimate before the survey has started taking data,
  and there is a range of forecasted values in the literature; for example,
  $\langle\gamma_{\rm int}^2\rangle^{1/2}=0.23$ \citep{las11,kir11}.  We
  tested using rms ellipticity of 0.26 with the template photo-zs, and found
  that the change affects primarily the fiducial constraints, degrading
  e.g.\ marginalized error in $w$ by a factor of $\sim 1.6$ (from 0.035 to
  0.055).  The overall degradation in the $\sigma_{68}$ of the distribution of
  $|\delta w|/\sigma(w)$ degrades by a factor of $\sim 1.9$ for the LSS cases
  and $\sim 1.6$ for the random equivalent cases.  Since we find that
  the intrinsic galaxy ellipticity primarily affects the fiducial cosmological
  parameter errors (i.e.\ $\sigma(w)$, rather than the systematic bias $\delta
  w$), we use it as a control parameter to vary our baseline
  cosmological parameter error assumptions\footnote{Note, it would not be hard
    to come up with other ways to improve the fiducial constraints, such as
    adding other 2-pt correlations to the analysis, or including
    magnification.  Conversely, one could add intrinsic alignments and other
    sources of errors to degrade the constraints.}.  Henceforth, we adopt
  $\langle\gamma_{\rm int}^2\rangle^{1/2}=0.16$ as the optimistic case for the
  dark energy fiducial errors (which leads to more challenging follow-up
  requirements), and $\langle\gamma_{\rm int}^2\rangle^{1/2}=0.26$ as the
  pessimistic error case (which leads to more relaxed requirements).
  Unless mentioned otherwise, results assume the former, optimistic case.

\subsubsection{Dependence on redshift range}\label{sec:discz}
After the completion of the paper, we obtained a newer version of the 
DES simulations that reached $z=2$. 
We found that the redshift range $1.35-2$ only comprised about $6.5\%$
of the sample and had little impact on the results despite the significantly 
worse photo-zs for galaxies in that range. 
Fractional biases degrade by $10\%$, an effect driven primarily by the 
improvement in fiducial constraints - which assume perfect photo-zs.

\subsubsection{Dependence on number of tomographic bins}\label{sec:discb}

We have adopted a rather aggressive redshift slicing as our baseline case,
assuming 20 tomographic redshift bins distributed in the $0<z<1.35$
range. 
We expect that with fewer redshift slices, photo-z errors will be less 
pronounced while the statistical errors will increase slightly; and thus 
that the spectroscopic follow-up requirements derived in this paper will be somewhat
relaxed. This expectation is backed up by numerical checks that we now describe. 

In addition to $B=20$, we also consider cases of $B=5$, 10, 15, 30 and 40
tomographic bins using alternately the template and polynomial photo-z
methods. We find that the dependence of biases in cosmological constraints on
the number of bins is rather weak. As $B$ increases from 5 to 20, the bias in
the dark energy equation of state {\it decreases} by $\sim 30\%$ and converges
at this point, not increasing appreciably for higher $B$ (reflecting the fact
that such small-redshift-scale fluctuations are not degenerate with
cosmological information). 
Moreover, as $B$ increases from 5 to 20, the {\it statistical} errors on $w$
decrease by 10\%, and drop a further $\sim 10\%$ as $B$ is increased to
40. Therefore the bias-to-error ratio decreases by a total of $\sim 20\%$ up
to $B=20$ but then increases by $\sim 10\%$ for $B=20\rightarrow 40$.  Given
these unremarkable dependencies for such a wide range of $B$, and the fact
that higher $B$ implies more stringent requirements, we conclude that 20
tomographic bins is indeed a good representative choice for the calculations in
this paper.

\section{Discussion: Can things be improved?}\label{sec:improv}

In this section, we discuss possibilities for reducing the impact of sample
variance.  In Sec.~\ref{sec:improvnow}, we present tests we have performed and
in Sec.~\ref{sec:improvfut} we discuss other possibilites that should be
explored.

\subsection{Performed tests}\label{sec:improvnow}

\begin{itemize}

\item {\it Culling:} We used the width of the \pofzw\ as a criterion to
  identify catastrophic photo-zs.  We removed all galaxies for which
  $\sigma(p(z))\geq 0.15$, which culled $10\%$ of the galaxies in our
  simulation.  The impact of this selection is summarized in Table
  \ref{tab:wbiascul}.  The scatter in the photo-zs improved by $13\%$ and
  $15\%$ for the template and polynomial methods, respectively, and the mean
  \pofzw\ width improved $10\%$.  The width of the fractional $w$ bias
  distribution, as described by $\sigma_{68}$ improved by $6$ and $11\%$ for
  the polynomial and \pofzw\ techniques, respectively, but only improved the
  template estimator results by the negligible $3\%$.

  We also tried to perform the culling using an error estimate from
  the template-fitting code\footnote{The error estimate we use is the
  difference between the {\tt Z\_BEST68\_HIGH} and {\tt Z\_BEST68\_LOW} 
  outputs of the {\it LePhare} code.}.  The results are in the entry
  Template*, in Table \ref{tab:wbiascul}.  We see that the template
  error estimation was less efficient than the \pofzw\ width for
  improving the photo-z scatter.  With the same fraction of objects
  removed, the mean scatter improved by only $8\%$ compared to $13\%$
  when the \pofzw\ width was used.  In addition, the culling actually
  resulted in worsening of the bias in $w$, despite an improvement in
  the overall cosmological parameter fit measured by the improvement
  in the median $\chisq$.
  
  The  conclusion is that culling of outliers does not seem to be a very efficient
  way to improve the bias due to photo-z calibration even when it works reasonably 
  well in improving the mean photo-z scatter.  

\item $\pzp$: If the true redshift distribution of the photometric sample is
  known somehow (e.g. using cross-correlation techniques \citep{Newman}, 
  or from theoretical priors), then one can use it to improve
  results.  As discussed in Sec.~\ref{sec:varvar}, the quantity $\pzp$ is much
  less sensitive to sample variance than $\pzs$.  
  If $N(z_s)$ for the photometric sample is known, we use the fact that
\begin{equation}
P(z^i_s|z^j_p)=P(z^j_p|z^i_s)\frac{N^i_s}{N^j_p}
\end{equation}
to estimate $\pzp$ from $\pzs$.  Table \ref{tab:wbiassmth} shows the
improvement in the statistics of the dark energy equation of state bias.  For
the $6 \degs$ case, we see from the last column that the statistics
from template-fitting and \pofzw\ methods improve by a factor of $\sim 5$
relative to the fiducial results shown in Table \ref{tab:wbias}; this
corresponds to 25 times smaller follow-up samples needed to achieve the same
calibration!  Improvements for the $1 \degs$  are not as pronounced, 
but are still substantial.  
These results are idealized, because the redshift distribution
is assumed to be perfectly known.  
How well does $N(z_s)$ neeed to be known for this technique to 
be useful is an open question.

If one uses a \pofz\ estimator (from any algorithm), the \pofz s can be
corrected using the improved $\pzp$.  The ability to correct the redshift
estimates is only possible for \pofz\ estimators but not for single-value
photo-zs.

\end{itemize}

\subsection{Other possible improvements}\label{sec:improvfut}

In this section we briefly describe potentially interesting techniques 
to reduce the spectroscopic follow-up requirements, but that go beyond 
the scope of this paper.

\begin{itemize}
\item Smoothing, fitting and deconvolution. With enough theoretical 
priors, one may use assumptions about smoothness or a functional form 
of the overall redshift distribution to fit the weights estimate
of the redshift distribution.
Alternatively, since the redshift sample variance is due to the projection 
along the line-of-sight of the linear power spectrum, one can perhaps use 
Fourier techniques to deconvolve the large-scale-structure from the redshift 
distribution estimates.

\item Repeat observations. The use of repeat photometric observations would 
help reduce the shot-noise component of the photo-z training procedure.
Unfortunately, the sample variance would not be affected.
The reduction of such noise might be relevant to help stabilize 
deconvolution techniques.

\end{itemize}

\begin{table}
\begin{center}
\begin{tabular}{c|ccc|ccc}
\hline\hline   \multicolumn{7}{c}{\rule[-2mm]{-3mm}{6mm} \bf{Bias in w (with culling)} }\\
\hline\hline   \multicolumn{7}{c}{\rule[-2mm]{-3mm}{6mm}  $\mathbf{6 \degsb}$}\\
\hline          Technique & \rule[-2mm]{-3mm}{6mm} \dwmean& \rule[-2mm]{-3mm}{6mm} $\sigma_{68}$ & \rule[-2mm]{-3mm}{6mm} $\chisqm$ & \rule[-2mm]{-3mm}{6mm} $R(\sigma_z)$    & \rule[-2mm]{-3mm}{6mm} $R(\sigma_{68})$ & \rule[-2mm]{-3mm}{6mm} $R(\chisqm)$  \\\hline
\rule[-2mm]{-3mm}{6mm}Template    & 0.01  & 2.48 & 3.20   &0.87 &0.97 &1.02 \\
\rule[-2mm]{-3mm}{6mm}Template*   & -0.06 & 2.90 & 2.59   &0.92 &1.13 &0.82 \\
\rule[-2mm]{-3mm}{6mm}Polynomial  & -0.17 & 1.44 & 1.75   &0.85 &0.94 &0.86 \\
\rule[-2mm]{-3mm}{6mm}\pofzw        & 0.03  & 2.08 & 1.94   &0.90 &0.89 &0.75 \\\hline\hline
\end{tabular}
\caption{Mean and $\sigma_{68}$ scatter of the fractional bias in $w$ for the
  different techniques, assuming patches of area 6, 1, $1/4 \degs$ for
  training and calibration or a random subsample with the same number of
  galaxies.  The $\chisqm$ column indicates the median $\chisq $ of the fit
  over all cosmological parameters.  In this Table, $10\%$ of the galaxies
  were removed based on \pofzw\ width.  The $R(\sigma_{z})$ shows the ratio of
  the photo-z scatters (or the \pofzw width) of results on this Table to the
  corresponding value in Table \ref{tab:wbias}.  The $R(\sigma_{68})$ shows
  the ratio of the $\sigma_{68}$ used in this Table, to the corresponding
  value in Table \ref{tab:wbias}, assuming the same fiducial statistical
  constraint for both cases.  As a result, this ratio compares the change in
  total bias, not fractional.
  To get the change in fractional bias one should note that the culling degrades the 
  statistical constraints on $w$ by $6\%$.
  }
\label{tab:wbiascul}
\end{center}
\end{table}

\begin{table}
\begin{center}
\begin{tabular}{c|ccc|cc}
\hline\hline   \multicolumn{6}{c}{\rule[-2mm]{-2mm}{6mm} \bf{Bias in w (with $\pzp$)} }\\
\hline\hline   \multicolumn{6}{c}{\rule[-2mm]{-2mm}{6mm}  $\mathbf{6 \degsb}$}\\
\hline   Technique & \rule[-2mm]{-2mm}{6mm} Mean& \rule[-2mm]{-2mm}{6mm} $\sigma_{68}$ & \rule[-2mm]{-2mm}{6mm} $\chisqm$    & \rule[-2mm]{-2mm}{6mm} $R(\sigma_{68})$ & \rule[-2mm]{-2mm}{6mm} $R(\chisqm)$  \\\hline 
\rule[-2mm]{-2mm}{6mm}Template    & -0.06 & 0.52 & 0.36&0.20 &0.11   \\
\rule[-2mm]{-2mm}{6mm}Polynomial  & -0.13 & 0.87 & 0.43&0.57 &0.21  \\
\rule[-2mm]{-2mm}{6mm}\pofzw        & -0.14 & 0.52 & 0.34&0.22 &0.13  \\\hline
\multicolumn{6}{c}{\rule[-2mm]{-2mm}{6mm}  $\mathbf{1 \degsb}$}\\\hline
\rule[-2mm]{-2mm}{6mm}Template    & -0.17 & 1.28 & 1.39 &0.34&0.19  \\
\rule[-2mm]{-2mm}{6mm}Polynomial  & -0.39 & 1.31 & 1.69 &0.44&0.36  \\
\rule[-2mm]{-2mm}{6mm}\pofzw        & -0.29 & 0.98 & 1.14 &0.25&0.13  \\\hline\hline
\end{tabular}
\caption{Mean and $\sigma_{68}$ scatter of the fractional bias in $w$ for the
  different techniques, assuming patches of area 6, 1, $1/4 \degs$ for
  training and calibration or a random subsample with the same number of
  galaxies.  The $\chisqm$ column indicates the median $\chisq $ of the fit
  over all cosmological parameters.  Results in this Table assume the true
  redshift distribution of the photometric sample was known, allowing us to
  use $\pzp$ instead of $\pzs$ as described in the text.  The $R(\sigma_{68})$
  shows the ratio of the $\sigma_{68}$ used in this Table, to the
  corresponding value in Table \ref{tab:wbias}.  }
\label{tab:wbiassmth}
\end{center}
\end{table}

\section{Guide for observing proposals}\label{sec:guide}
In this section we provide a guide for observers to determine what observing
requirements are needed for photo-z calibration given a specific telescope's
effective angular aperture, number of spectroscopic fibers and collecting
area.  Typically, calibration requirements have been represented in terms of
total number of galaxies.  We argue that calibration requirements should be
phrased in terms of variables more closely related to total observing time or
cost.  With this purpose in mind, we define the number of pointings, $\npoint$,
to be the product of the number of patches times the number of repeat
observations of each patch.  For constant collecting area, the number of
pointings is a direct measure of total observational time required.

The previous sections focused on calibration requirements from a single patch.
If independent patches are combined, the requirements decrease with the
square-root of the number of independent patches.  This square-root scaling
only applies exactly to the template-fitting method because it does not use a
training procedure. 
For simplicity, and because the previous results were rather insensitive to
photo-zs, we only use the template photo-zs in this section.
 
As an example, we consider the case of the Dark Energy Survey.
To reach reasonable spectroscopic completeness at the limiting magnitudes of
the DES requires very large telescopes.  We thus tune our guide to two of the
telescopes that will be available for the calibration: VIMOS-VLT and
IMACS-Magellan.  VLT is an 8-meter class telescope with angular aperture of
$250 \acmins$ (or about $1/16 \degs$). 
Magellan is a 6.5-meter class telescope with collecting area of $0.25 \degs$.  
We assume that in each observation, VLT and Magellan can observe about 
300-500 galaxies if a low-dispersion setting is used.
In real observations, the need to disperse the spectra in the focal plane 
reduces much of the available collecting area.  
This is not a random reduction, however.  
Roughly speaking, spectra cannot be at the edges of the focal plane so that there
is room left in the focal plane to disperse the spectra. 
The design of VLT already accounts for this, but for Magellan there is a
loss of up to a half of the total area.
To roughly cover the possibilities for existing telescopes of large angular aperture
we perform our tests assuming $1/4$, $1/8$ and $1/32 \degs$ fields-of-view.

Fig.~\ref{fig:survcalc} shows the number of independent patches that must be
observed as function of the number of galaxies per patch so that the photo-z
calibration leads to bias in $w$ that is smaller than the statistical error in
$w$ with $95\%$ probability.  One can see that, for fixed number of galaxies
per patch,
the larger the telescope, the smaller the number of independent patches that
need to be observed.  Hence, assuming equal throughput and same number of
available fibers, a telescope such as Magellan is more efficient than VLT for
spectroscopic calibration.  For reference, we also show the results assuming
the full $1/4 \degs$ field-of-view of Magellan is available for spectroscopy.
For the case of the $1/4 \degs$ collecting area, if the telescope can observe
400 galaxies at once, then about 140 independent patches -- or a total of
$5.6\times 10^4$ galaxies -- would be needed to ensure, with $95$ probability
that the bias in the equation of state is less than the statistical error
(i.e.\ bias/error $\leq 1.0$).  The requirement increases to about 150 and 180
patches for effective angular apertures of $1/8 \degs$ and $1/32 \degs$.
The requirement for VLT would be about 165 patches (not shown).

The contours in Fig.~\ref{fig:survcalc} were constructed by varying the mean 
fraction of galaxies that
are sampled from each patch.  The right tip of each contour line corresponds
to using $100\%$ of the galaxies in a patch.  For a fixed angular aperture,
the total number of galaxies required decreases with decreasing sampling
fraction.  By sampling less galaxies per patch one more efficiently beats down
the sample variance, up to the point where shot noise dominates.  The total
number of galaxies required can never be smaller than the requirements from
shot-noise only estimates.  In our case, this is about $4\times 10^4$
galaxies.
The upturn in the contours at low sampling fraction indicates the shot-noise domination
regime, at which point reducing the number of galaxies per patch yields no benefit.

How does one use Fig.~\ref{fig:survcalc} to deduce more stringent requirements
on dark energy parameter biases, or implement different survey assumptions?
The distribution of fractional bias in $w$ is roughly Gaussian, hence to get
$N$-$\sigma$ requirements on the bias, one can simply multiply the
$2$-$\sigma$ requirement plotted by $N/2$.  For example, the requirement of
keeping the bias/error less than $1.0$ at $2$-$\sigma$ roughly implies that
the bias is less than $0.5$ at $1$-$\sigma$.  One can use a square-root
scaling to deduce more stringent requirements; for example, if one would like
the bias/error in $w$ to be less than $0.25$ at $1$-$\sigma$, then the number
of independent patches required increases by four.  Because the effect of the
independent number of patches is only a square-root, it is well worth
investigating techniques that decrease the sensitivity to the sample variance.
For example, as we saw in Sec.~\ref{sec:improv}, if the redshift distribution
of the phometric sample could be perfectly known, the calibration biases would
decrease by factors of up to 5 which would decrease the number of patches
required for photo-z calibration by more than a factor of 25!  

  Finally, recall that usage of a more realistic intrinsic galaxy ellipticity
  of 0.26 increases the fiducial $w$ error by a factor of 1.6 (from 0.035 to
  0.055) and leaves the biases in $w$ largely unaffected, resulting in the
  decreased follow-up requirements by a factor of $\sim 3.5$ in the number of
  patches required. Nevertheless, we think that usage of the smaller value of
  the intrinsic rms ellipticity used throughout is preferred, given that the
  fractional biases $\delta w/\sigma(w)$ could be larger than expected. This
  could happen in two ways: either the fiducial error $\sigma(w)$ could be
  improved by other weak lensing techniques (3-point function, other
  cross-correlations, etc), or additional systematics 
  might increase the bias $\delta w$.  We therefore erred on the side of being
  conservative in terms of the spectroscopic follow-up requirements, and
  adopted $\langle\gamma_{\rm int}^2\rangle^{1/2}=0.16$, or
  $\sigma(w)=0.035$. Our best current understanding is that
 only three kinds of systematics would increase spectroscopic follow-up
requirements: non-random spectroscopic failures, imperfect
star-galaxy separation, and variability in observing conditions. Other
systematics would likely only cause a degradation in the
  fiducial cosmological parameter constraints, thereby decreasing follow-up
requirements.

The time required for completing observations depends on the requirements on
spectroscopic completeness.  If we assume that a completeness level comparable
to that of the VIMOS-VLT Deep Survey\footnote{\url{
    http://cesam.oamp.fr/vvdsproject/}} (VVDS) is
sufficient\footnote{The VVDS-DEEPS survey obtained redshifts for about
    $44\%$ of their sample with confidence above $91-97\%$, and of these,
    about $22\%$ had confidence of $99\%$ \citep{lef05} }, then two patches
of sky can be covered per night using VLT or Magellan, if a single pointing is
required per patch.  In the absence of spectroscopic failures, the ideal
strategy is clearly to use a single pointing per patch to beat down sample
variance as fast as possible.  However, spectroscopic failures typically
cannot be ignored, which makes it harder to determine the optimal observing
strategy.  The key difficulty is that spectroscopic failure rates vary
strongly with galaxy type, which implies that different observing times are
needed for different types of galaxies to yield reliable redshifts.  In
addition, for a fixed galaxy type, there is a broad distribution of intrinsic
luminosities.  An optimized survey would, at the very least, require a
carefully weighted target selection function to ensure the final spectroscopic
sample is a representative subsample of the full photometric survey.  At best,
the ideal survey would combine several telescopes, each optimized for a
certain depth and galaxy population.  For example, planned surveys such as
BigBOSS\footnote{\url{http://bigboss.lbl.gov/}} and
DESpec\footnote{\url{http://eag.fnal.gov/DESpec/Home.html}} will have very
wide fields of view and be able to obtain several thousand spectra per
pointing.  An interesting strategy would be to use these telescopes -- perhaps
with massive coaddition of images -- to obtain a large sample to depths
slightly brighter than $i\simeq 24$, for galaxy types with more easily
detectable spectra.  This way, an 8-m class telescope could concentrate
exclusively on the very faintest galaxies.

In a forthcoming follow-up to this paper, we incorporate a simulated spectroscopic 
pipeline to our analysis to determine the levels of spectroscopic completeness that 
are required for dark energy studies.

\begin{figure*}
  \includegraphics[scale=0.5,angle=-90]{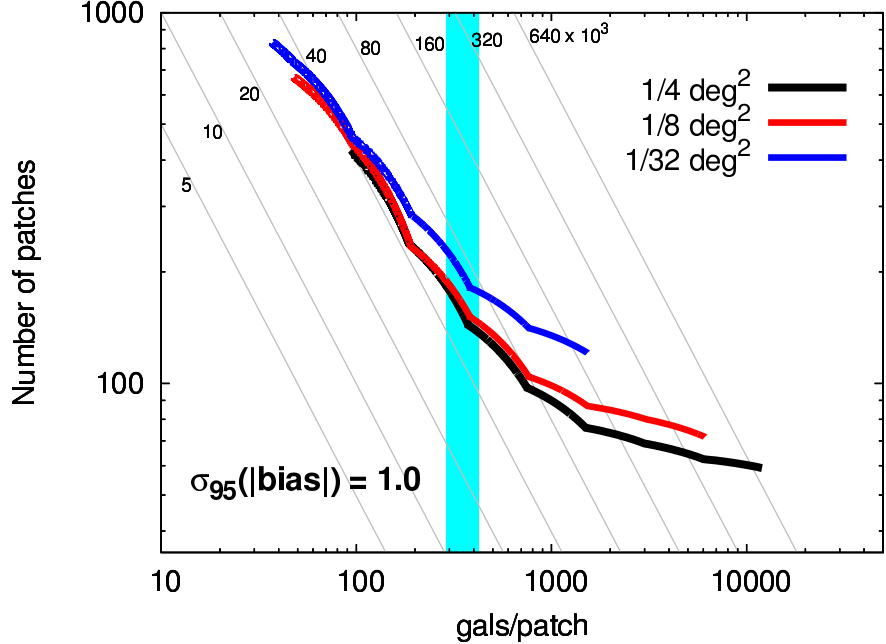}\hspace{0.5cm}
\caption{Relation between number of independent patches and
  galaxies observed per patch so that the calibration bias will yield a
  bias/error ratio in $w$ that is less than 1.0 with $95\%$ probability. We
  consider three different telescope apertures based on capabilities of
  existing telescopes: $1/4 \degs$ (solid black), $1/8 \degs$ (solid red) and
  $1/32 \degs$ (or $112.5 \acmins$; blue). 
  The first two scenarios correspond to the optimistic and pessimistic assumptions
  about the effective observing area of Magellan.
  The VIMOS-VLT instrument could observe about $1/16 \degs$. 
  The diagonal light gray lines indicate contours of fixed
  total number of galaxies, while the vertical band indicates typical number
  of galaxies per observed patch possible with a single pointing of Magellan
  or VLT.  For a fixed number of galaxies per patch, the total number of
  patches required is higher for a smaller patch area in order to compensate
  for the increased sample variance per patch. Similarly, if the survey can
  observe more galaxies in each patch, then the total number of patches
  obviously decreases since fewer patches will be required to calibrate the
  shot noise, at the expense of increasing the total number of galaxies required.
}
\label{fig:survcalc}
\end{figure*}


\section{Conclusions}\label{sec:concl}

We used cosmological N-body simulations populated with galaxies with
DES photometry to investigate the impact of shot-noise and sample
variance in the spectroscopic observations necessary to train the
photo-zs and calibrate their error distributions.  Our conclusions are
as follows:

\begin{itemize}
\item For typical spectroscopic surveys, sample variance is much larger than shot noise.

\item Sample variance affects the spectroscopic properties more strongly than
  photometric properties.  Consequently, the error distribution $\pzs$ is much
  more sensitive to sample variance than $\pzp$.  Unfortunately, for
  cosmological analysis $\pzs$ is the error distribution that we have to use,
  which results in calibration requirements that are quite demanding.  If the
  overall distribution of the photometric sample is known somehow, e.g.\ using
  cross-calibration techniques, then one can estimate $\pzs$ from $\pzp$,
  which can reduce follow-up requirements by more than an order of magnitude.
  In addition, if one uses \pofz s instead of single-number photo-z estimates,
  the improved $\pzp$ estimate can be used to correct and improve the \pofz s.

\item The use of the same spectroscopic sample to train photo-zs and calibrate
  the photo-z error distribution does not introduce additional cosmological
  biases.
  In addition, the scatter in the photo-zs is, on average, not degraded by 
  sample variance.

\item  For small training sets the \pofzw\ method is the most affected by sample 
variance because it is a pure density estimator (cf. Fig.~\ref{fig:nzp}).  
Conversely, the \pofzw\ estimate is  the only unbiased method in the sense 
that, for large enough training, it  recovers the true redshift distribution of the photometric sample.

\item Biases in the dark energy equation of state obtained from the different
  photo-z methods are highly correlated for sample-variance-dominated
  calibration samples, suggesting that a simple combination of photo-z methods
  cannot reduce the biases.  Conversely, for shot-noise dominated calibration
  samples, biases are largely uncorrelated.

\item Culling of catastrophic outliers is not very effective at
  reducing calibration requirements, with the decrease in the bias in $w$
  being comparable to degradation of the statistical errors due to the
  reduction of the sample size.
  
\item We provide a guide to observing proposals of spectroscopic samples
  directed towards the calibration of photo-zs for the DES.  We focus on
  Magellan and VLT, the two telescopes best suited for DES calibration.  To
  reduce sample variance effects one should spread the observations to as many
  patches as possible, using as many spectroscopic fibers as possible in each
  observation.  We find that VLT and Magellan would need about 165 and 150
  patches respectively in order to ensure, with $95\%$ probability, that the
  photoz-calibration induced bias in $w$ does not dominate its statistical
  error.  This estimate assumes that 400 galaxies can be observed per patch.
  If a VVDS-level of completeness is sufficient, these observations would
  require about 85 and 75 nights of observation for VLT and Magellan,
  respectively, assuming the optimistic fiducial uncertainty of $\sigma(w) =
  0.035$.  For a more pessimistic fiducial error $\sigma(w)= 0.055$, the
  requirements decrease by a factor of about 3.5.  Nevertheless, the former
  number may be more useful as a guideline, since the overall requirements
  might be increased by including the type incompleteness and spectroscopic
  redshift failures, something that we will fully investigate in a forthcoming
  companion paper.

\end{itemize}

\section*{Acknowledgements}

CC would like to thank Joerg Dietrich, Huan Lin, Anja von der Linden and Jeff
Newman for discussions about spectroscopic surveys and Stephanie Jouvel for
help with the {\it Le Phare} code.  We thank Gary Bernstein, Joanne Cohn,
Martin White, and an anonymous referee for very useful comments.  We
would like to thank the Kavli Institute for Theoretical Physics in Santa
Barbara where some of this work was carried out.  MTB and RHW would like to
thank their collaborators of the LasDamas project for use of their simulation
data.  CC and DH are supported by the DOE OJI grant under contract
DE-FG02-95ER40899. CC is also supported by a Kavli Fellowship at Stanford
University. DH is additionally supported by NSF under contract AST-0807564,
and NASA under contract NNX09AC89G.  RHW recieved support from the
U.S. Department of Energy under contract number DE-AC02-76SF00515.  MTB was
supported by Stanford University and the Swiss National Science Foundation
under contract 2000 124835/1.  This research was supported in part by the
National Science Foundation under Grant No. PHY05-51164.

\appendix
\section{The simulations} \label{app:sims}

The simulated galaxy catalog used for the present work was generated
using the Adding Density Determined GAlaxies to Lightcone Simulations
(ADDGALS) algorithm \citep{wechsler-addgals,busha-addgals}.  This
algorithm attaches synthetic galaxies to dark matter particles in a
lightcone output from a dark matter N-body simulation.  The model
is desgined to match the luminosities, colors, and clustering
properties of galaxies.

The simulation used here was based on a single ``Carmen'' simulation
from the LasDamas project \citep{mcbride11}.  This simulation was run
with the publicly available Gadget-2 code and modeled a flat $\Lambda
CDM$ cosmology with $\Omega_m = 0.25$ and $\sigma_8= 0.8$ in a
1Gpc/$h$ box with $1120^3$ particles.  The lightcone output necessary
for the ADDGALS algorithm was created by pasting together 33 snapshots
in the redshift range $z = 0-1.33$.  This results in a 220 sq degree
lightcone whose orientation was selected such that there are no
particle replications in the inner $~\sim 100$ sq.~deg. and minimal
replications in the outer regions.

The ADDGALS algorithm used to create the galaxy distribution consists
of two steps: galaxies based on an input luminosity function are first
assigned to particles in the simulated lightcone, after which
multi-band photometry is added to each galaxy using a training set of
observed galaxies.  For the first step, we begin by defining the
relation $P(\delta_{dm}|M_r, z)$ --- the probability that a galaxy
with magnitude $M_r$ a redshift $z$ resides in a region with local
density $\delta_{dm}$, defined as the radius of a sphere containing
$1.8 \times 10^{13} \hinv\Msun$ of dark matter.  This relation can be
tuned to reproduce the luminosity-dependent galaxy 2-point function by
using a much higher resolution simulation combined with the technique
known as subhalo abundance matching.  This is an algorithm for
populating very high resolution dark matter simulations with galaxies
based on halo and subhalo properties that accurately reproduces
properties of the observed galaxy clustering \citep{conroy06,
  wetzel10, behroozi10, busha11}.  The relationship
$P(\delta_{dm}|M_r,z)$ can be measured directly from the resulting
catalog.  Once this probability relation has been defined, galaxies
are added to the simulation by integrating a (redshift dependent)
$r$-band luminosity function to generate a list of galaxies with
magnitudes and redshifts, selecting a $\delta_{dm}$ for each galaxy by
drawing from the $P(\delta_{dm}|M_r, z)$ distribution, and attaching
it to a simulated dark matter particle with the appropriate
$\delta_{dm}$ and redshift.  The advantage of ADDGALS over other
commonly used approaches based on the dark matter halos is the ability
to produce significantly deeper catalogs using simulations of only
modest size.  When applied to the present simulation, we populate
galaxies as dim as $M_r \approx -16$, compared with the $M_r \approx
-21$ completeness limit for a standard halo occupation (HOD) approach.

While the above algorithm accurately reproduces the distribution of
satellite galaxies, central objects require explicit information about
the mass of their host halos.  Thus, for halos larger than $5 \times
10^{12}\hinv\Msun$, we assign central galaxies using the explicit
mass-luminosity relation determined from our calibration catalog.  We
also measure $\delta_{dm}$ for each halos, which is used to draw a
galaxy from the integrated luminosity function with the appropriate
magnitude and density to place at the center.

For the galaxy assignment algorithm, we choose a luminosity function
that is similar to the SDSS luminosity function as measured in
\cite{bla03}, but evolves in such a way as to reproduce the higher
redshift observations of the NDWFS and DEEP2 observations.  We use a
Schechter Function with $\phi* = 1/81\times 10^{-2z/3}$, $M_* = -20.34
+ 3.5*(a-0.91)$, and $\alpha = -1.03$, where $a$ is the cosmological
expansion factor.

Once the galaxy positions have been assigned, photometric properties
are added.  We begin with a training set of spectroscopic galaxies and
the simulated set of galaxies with $r$-band magnitudes generated
earlier.  For each galaxy in both the training set and simulation we
measure $\Delta_5$, the distance to the 5th nearest galaxy on the sky
in a redshift bin.  Each simulated galaxy is then assigned an SED
based on drawing a random training-set galaxy with the appropriate
magnitude and local density, k-correcting to the appropriate redshift,
and projecting onto the desired filters.
The kcorrections and projections are performed using the Kcorrect code 
\citep{bla03}. The construction of the SEDs in Kcorrect is described
in \cite{bla07}.

Differences between the training set and simulated galaxy sample
complicate the process of color-assignment.  In order to compile a
sufficiently large training set, we use a magnitude-limited sample of
SDSS spectroscopic galaxies brighter than $m_r = 17.77$ with $z <
0.2$.  The simulated sample, on the other hand, is a volume-limited
sample, spanning a broader redshift range.  When measuring $\Delta_5$
we restrict ourselves to neighbors brighter than $M_r = -19.7$ in the
simulation sample, while using all objects in the observational
catalog.  To mitigate differences in luminosity and redshift,
each galaxy is rank ordered according to its density in its redshift bin,
and require that objects be in the same percentile bin in each sample
rather than having the same the absolute value of $\Delta_5$.  This is
similar to the method used in \cite{cooper08}.

The final step for producing a realistic simulated catalog is the
application of photometric errors.  While the photometric errors
generated here are particular to DES, the algorithm can be generalized
for any survey.  For each galaxy, we add a noise term to the intrinsic
galaxy flux, where the noise is drawn from a Gaussian of width
\begin{equation}
{\rm noise} = \sqrt{t_e n_p n_s + f_{g,i} t_e}
\end{equation}
where $t_e$ is the exposure time, $n_p$ the number of pixels covered by a 
galaxy, $n_s$ the flux of the sky in a single detector pixel, and $f_{g,i}$ 
is the intrinsic flux of the galaxy.  Application of the above relation to 
objects from the SDSS catalog shows that it is able to faithfully reproduce 
the reported errors of the survey.

\bibliographystyle{mn2e_new}
\bibliography{paper}

\end{document}